\documentclass[longauth]{aa} % for the long lists of affiliations 
\usepackage{xspace}
\usepackage{deluxetable}   
\usepackage{graphicx}
\usepackage{color}
\usepackage{lineno}
\usepackage{graphics}
\usepackage{subcaption}
\usepackage[section]{placeins}
\usepackage{longtable}
\usepackage{graphicx}
\usepackage[export]{adjustbox}
\DeclareUnicodeCharacter{2212}{-}

\usepackage{txfonts}
\begin{document}

\title{HD207897 b: A dense sub-Neptune transiting a nearby and bright K-type star}

   \author{N. Heidari$^{*}$ \inst{1,2,3},
          I. Boisse \inst{\ref{lam}},
          J. Orell-Miquel \inst{\ref{spanish_team1},\ref{spanish_team2}},
          G.~H\'ebrard \inst{\ref{OHP},\ref{paris}},
          L.~Acu\~na \inst{\ref{lam}},
          N.~C.~Hara \inst{\ref{geneva}},
          J.~Lillo-Box \inst{\ref{jorge_image}},
          J.~D.\ Eastman \inst{\ref{jason}},
          L.~Arnold \inst{\ref{OHP},\ref{CFHT}},
          N.~Astudillo-Defru \inst{\ref{chile}},
          V.~Adibekyan \inst{\ref{santos1},\ref{santos2}},
          A. Bieryla \inst{\ref{jason}},
          X.~Bonfils \inst{\ref{gronobl}},
          F. Bouchy \inst{\ref{geneva}},
          T.~Barclay \inst{\ref{barcley2},\ref{barcley1}},
          C. E.~Brasseur \inst{\ref{Brasseur}},
          S.~Borgniet \inst{\ref{kevin}},
          V.~ Bourrier \inst{\ref{geneva}},
          L.~Buchhave \inst{\ref{danmark}},
          A.~Behmard \inst{\ref{NSF},\ref{Behmard}},
          C.~Beard \inst{\ref{Beard}},
          N. M~.Batalha \inst{\ref{Batalha}},
          B.Courcol \inst{\ref{geneva}},
          P.~Cort\'es-Zuleta \inst{\ref{lam}},
          K.~Collins \inst{\ref{jason}},
          A.~Carmona \inst{\ref{gronobl}},
          I. J. M. Crossfield \inst{\ref{Crossfield}},
          A.~ Chontos \inst{\ref{NSF},\ref{Chontos}},
          X.~Delfosse \inst{\ref{gronobl}},
          S.~Dalal \inst{\ref{paris}},
          M. Deleuil \inst{\ref{lam}},
          O. D. S. Demangeon \inst{\ref{santos1},\ref{santos2}},
          R.~ F. D\'iaz \inst{\ref{diaz}},
          X.~Dumusque \inst{\ref{geneva}},
          T.~ Daylan \inst{\ref{Daylan},\ref{kavli}},
          D.~Dragomir \inst{\ref{dragmir}},
          E.~Delgado~Mena \inst{\ref{santos1}} ,
          C.~Dressing \inst{\ref{Dressing}},
          F.~Dai \inst{\ref{Pasadena}},
          P. A.~Dalba \inst{\ref{california},\ref{nsf_postdoc}},
          D.~ Ehrenreich \inst{\ref{geneva}},
          T.~Forveille \inst{\ref{gronobl}},
          B.~Fulton \inst{\ref{NASA_california}},
          T.~ Fetherolf \inst{\ref{california}},
          G.~Gaisné \inst{\ref{gronobl}},
          S.~Giacalone \inst{\ref{Dressing}},
          N.~Riazi \inst{1},
          S.~Hoyer \inst{\ref{lam}},
          M.~J. Hobson \inst{\ref{melissa1},\ref{mellisa2}} 
          A. W.~Howard \inst{\ref{pasadena-california}},
          D.~Huber \inst{\ref{Chontos}},
          M. L.~Hill \inst{\ref{california}},
          L. A.\ Hirsch \inst{\ref{stanford}},
          H.~Isaacson \inst{\ref{Dressing},\ref{Australia}},
          J.~Jenkins \inst{\ref{Jenkins}},
          S. R.~Kane \inst{\ref{california}},
          F.~Kiefer \inst{\ref{paris},\ref{kevin}},
          R.~Luque \inst{\ref{spanish_team1},\ref{spanish_team2}},
          D.~ W. Latham \inst{\ref{jason}},
          J.~Lubin \inst{\ref{Beard}},
          T.~ Lopez \inst{\ref{lam}},
          O.~Mousis \inst{\ref{lam}},
          C.~Moutou \inst{\ref{Tuluse}},
          G.~Montagnier \inst{\ref{OHP}},
          L.~ Mignon \inst{\ref{gronobl}},
          A.~ Mayo \inst{\ref{Dressing}},
          T.~ Mo\v{c}nik \inst{\ref{Gemini}},
          J. M. A.~Murphy \inst{\ref{NSF},\ref{Batalha}},
          E. Palle \inst{\ref{spanish_team1},\ref{spanish_team2}},
          F. ~Pepe\inst{\ref{geneva}},
          E. A.~ Petigura \inst{\ref{Los Angeles}},
          J.~Rey \inst{\ref{rey}},
          G.~Ricker \inst{\ref{Daylan}},
          P.~Robertson \inst{\ref{Beard}},
          A.~Roy \inst{\ref{Brasseur},\ref{Johns Hopkins}},
          R. A.~Rubenzahl \inst{\ref{NSF},\ref{pasadena-california}},
          L. J.\ Rosenthal \inst{\ref{pasadena-california}},
          A.~Santerne \inst{\ref{lam}},
          N.~ C. Santos \inst{\ref{santos1},\ref{santos2}},
          S.~G.~Sousa \inst{\ref{santos1}},
          K.~G.~Stassun \inst{\ref{Stassun}},
          M.~ Stalport\inst{\ref{geneva}},
          N.~ Scarsdale \inst{\ref{Batalha}},
          P.~ A. Str{\o}m \inst{\ref{paris},\ref{wilson1},\ref{wilson2}},
          S.~Seager \inst{\ref{Daylan},\ref{seagar1},\ref{seagar2}},
          D.~ Segransan \inst{\ref{geneva}},
          P. ~Tenenbaum \inst{\ref{Tenenbaum}},
          R.~Tronsgaard \inst{\ref{danmark}},
          S.~ Udry \inst{\ref{geneva}},
          R.~ Vanderspek \inst{\ref{Daylan}},
          F.~ Vakili \inst{\ref{OCA}},
          J.~ Winn \inst{\ref{winn}},
          L. M. Weiss \inst{\ref{Chontos}}
           }
        
    \institute{Department of Physics, Shahid Beheshti University, Tehran, Iran.\\
             \email{neda.heidari@lam.fr}
    \and
         Laboratoire J.-L. Lagrange, Observatoire de la C\^ote d’Azur (OCA), Universite de Nice-Sophia Antipolis (UNS), CNRS, Campus Valrose, 06108 Nice Cedex 2, France.\label{OCA}
    \and
         Aix Marseille Univ, CNRS, CNES, LAM, Marseille, France.\label{lam}
    \and
         Instituto de Astrof\'isica de Canarias, 38205 La Laguna, Tenerife, Spain.\label{spanish_team1}
    \and
        Departamento de Astrof\'isica, Universidad de La Laguna, 38206 La Laguna, Tenerife, Spain.\label{spanish_team2}
    \and
       Observatoire de Haute-Provence, CNRS, Universit\'e d'Aix-Marseille, 04870 Saint-Michel-l'Observatoire, France \label{OHP}
    \and
       Institut d'astrophysique de Paris, UMR 7095 CNRS université pierre et marie curie, 98 bis, boulevard Arago,  75014, Paris \label{paris}
    \and
       Observatoire de Gen\`eve,  Universit\'e de Gen\`eve, Chemin Pegasi, 51, 1290 Sauverny, Switzerland\label{geneva}
    \and   
       Centro de Astrobiolog\'ia (CAB, CSIC-INTA), Depto. de Astrof\'isica, ESAC campus, 28692, Villanueva de la Ca\~nada (Madrid), Spain \label{jorge_image}   
    \and 
       Center for Astrophysics \textbar \ Harvard \& Smithsonian, 60 Garden St, Cambridge, MA 02138, USA \label{jason}
    \and
        Canada France Hawaii Telescope Corporation (CFHT), 65-1238 Mamalahoa Hwy, Kamuela HI 96743  USA \label{CFHT}
     \and
         Departamento de Matem\'atica y F\'isica Aplicadas, Universidad Cat\'olica de la Sant\'isima Concepci\'on, Alonso de Rivera 2850, Concepci\'on, Chile\label{chile}   
    \and
         Instituto de Astrof\'isica e Ci\^encias do Espa\c{c}o, Universidade do Porto, CAUP, Rua das Estrelas, 4150-762 Porto, Portugal.\label{santos1}
    \and     
         Departamento de F\'isica e Astronomia, Faculdade de Ci\^encias, Universidade do Porto, Rua do Campo Alegre, 4169-007 Porto, Portugal.\label{santos2}
    \and
       Univ. Grenoble Alpes, CNRS, IPAG, 38000 Grenoble, France\label{gronobl}     
    \and
       NASA Goddard Space Flight Center, 8800 Greenbelt Rd, Greenbelt, MD 20771, USA \label{barcley1}
    \and
        University of Maryland, Baltimore County, 1000 Hilltop Cir, Baltimore, MD 21250, USA \label{barcley2}
    \and
    Space Telescope Science Institute, 3700 San Martin Dr, Baltimore, MD 21218, USA \label{Brasseur}
    \and
       NSF Graduate Research Fellow \label{NSF}
    \and
       Division of Geological and Planetary Science, California Institute of Technology, Pasadena, CA 91125, USA \label{Behmard}
    \and
        Department of Physics \& Astronomy, University of California Irvine, Irvine, CA 92697, USA \label{Beard}
    \and   
        Department of Astronomy and Astrophysics, University of California, Santa Cruz, CA 95060, USA \label{Batalha}
    \and
        Department of Physics \& Astronomy, University of Kansas, 1082 Malott, 1251 Wescoe Hall Dr., Lawrence, KS 66045, USA \label{Crossfield}   
    \and
        Institute for Astronomy, University of Hawai`i, 2680 Woodlawn Drive, Honolulu, HI 96822, USA\label{Chontos}
    \and 
    International Center for Advanced Studies (ICAS) and ICIFI (CONICET), ECyT-UNSAM, Campus Miguelete, 25 de Mayo y Francia, (1650) Buenos Aires, Argentina.\label{diaz}
    \and
    Department of Physics, and Kavli Institute for Astrophysics and Space Research, Massachusetts Institute of Technology, Cambridge, MA 02139, USA \label{Daylan}
    \and
    Department of Physics and Astronomy, University of New Mexico, 1919 Lomas Blvd NE, Albuquerque, NM 87131, USA \label{dragmir}
    \and
     Kavli Fellow \label{kavli}
    \and
    Department of Astronomy, University of California Berkeley, Berkeley, CA, 94720, USA \label{Dressing}
    \and
    Division of Geological and Planetary Sciences 1200 E California Blvd, Pasadena, CA, 91125, USA \label{Pasadena}
    \and
     Department of Earth and Planetary Sciences, University of California, Riverside, CA 92521, USA\label{california}
    \and
       NSF Astronomy and Astrophysics Postdoctoral Fellow\label{nsf_postdoc}
    \and
    NASA Exoplanet Science Institute/Caltech-IPAC, MC 314-6, 1200 E. California Blvd., Pasadena, CA 91125, USA\label{NASA_california}
    \and
     Department of Physics and Astronomy, University of California, Riverside, CA 92521, USA \label{california_physics}
    \and
     Millennium Institute for Astrophysics, Chile \label{melissa1}
    \and
     Instituto de Astrofísica, Facultad de Física, Pontificia Universidad Católica de Chile, Chile \label{mellisa2}
    \and
    Department of Astronomy, California Institute of Technology, Pasadena, CA 91125, USA \label{pasadena-california}
    \and
        Kavli Institute for Particle Astrophysics and Cosmology, Stanford University, Stanford, CA 94305, USA \label{stanford}
    \and    
        Centre for Astrophysics, University of Southern Queensland, Toowoomba, QLD, Australia \label{Australia}
    \and
    Department of Physics \& Astronomy, University of California Los Angeles, Los Angeles, CA 90095, USA \label{Los Angeles}
    \and
    Las Campanas Observatory, Carnegie Institution of Washington, Colina el Pino, Casilla 601 La Serena, Chile \label{rey}
    \and
       Department of Physics and Astronomy, Johns Hopkins University, 3400 N Charles St, Baltimore, MD 21218, USA\label{Johns Hopkins}
    \and
       Gemini Observatory/NSF's NOIRLab, 670 N. A'ohoku Place, Hilo, HI 96720, USA \label{Gemini}
    \and
       LESIA, Observatoire de Paris, Université PSL, CNRS, Sorbonne Université, Université de Paris, 5 place Jules Janssen, 92195, Meudon, France \label{kevin}
    \and
     Univ. de Toulouse, CNRS, IRAP, 14 avenue Belin, 31400 Toulouse, France \label{Tuluse}
      \and
      Vanderbilt University, Department of Physics \& Astronomy, 6301 Stevenson Center Ln., Nashville, TN 37235, USA \label{Stassun}
      \and
      SETI Institute/NASA Ames Research Center, USA\label{Tenenbaum}
      \and
      DTU Space, National Space Institute, Technical University of Denmark, Elektrovej 328, DK-2800 Kgs. Lyngby, Denmark \label{danmark}
      \and
       Department of Astrophysical Sciences, Princeton University, 4Ivy Lane, Princeton, NJ 08544, USA\label{winn}
       \and
        NASA Ames Research Center, Moffett Field, CA, 94035, USA \label{Jenkins}
        \and
        Department of Physics, University of Warwick, Coventry, CV4 7AL, UK\label{wilson1}
        \and
        Centre for Exoplanets and Habitability, University of Warwick, Gibbet Hill Road, Coventry, CV4 7AL, UK \label{wilson2}
        \and
        Department of Earth, Atmospheric and Planetary Sciences, Massachusetts Institute of Technology, Cambridge, MA 02139, USA \label{seagar1}
        \and
        Department of Aeronautics and Astronautics, MIT, 77 Massachusetts Avenue, Cambridge, MA 02139, USA \label{seagar2}
       }

   \date{Received XX, 2020; accepted XX, 2020}

\abstract
{We present the discovery and characterization of a transiting sub-Neptune orbiting with a 16.20-day period around a nearby (28 pc) and bright (V= 8.37) K0V star HD207897 (TOI-1611). This discovery is based on photometric measurements from the Transiting Exoplanet Survey Satellite (TESS) mission and radial velocity (RV) observations from the SOPHIE, Automated Planet Finder (APF) and HIRES high precision spectrographs. We used EXOFASTv2 for simultaneously modeling the parameters of the planet and its host star, combining photometric and RV data to determine the planetary system parameters. We show that the planet has a radius of $2.50\pm0.08$  R$_{\mathrm{E}}$ and a mass of either $14.4\pm 1.6$  M$_{\mathrm{E}}$ or $15.9\pm1.6$ M$_{\mathrm{E}}$ with nearly equal probability; the two solutions correspond to two possibilities for the stellar activity period. Hence, the density is either $ 5.1\pm0.7$ g cm$^{-3}$ or $5.5^{+0.8}_{-0.7}$ g cm$^{-3}$, making it one of the relatively rare dense sub-Neptunes. The existence of such a dense planet at only 0.12 AU from its host star is unusual in the currently observed sub-Neptune (2 < R$_{\mathrm{E}}$ < 4) population. The most likely scenario is that this planet has migrated to its current position.}
  
\titlerunning{HD207897 b: A dense sub-Neptune transiting a nearby and brightK-type star}
\authorrunning{N. Heidari et al.}
  
  \maketitle
%
%-------------------------------------------------------------------

\section{Introduction}

The brightness of more than 200,000 stars has been monitored by TESS \citep{ricker2015jatis} with 2 minutes cadence during its two-year primary mission. The observed stars are closer and brighter (typically 30-100 times brighter) than the stars \emph{Kepler} surveyed. This offers us a unique opportunity for furthering our knowledge in planetary science with follow-up observations: those from ground-based high precision spectrographs to confirm the planetary nature and mass measurement, which with the radius allows us to determine the bulk composition of planets; and ground and space-based observations to provide atmospheric characterization, for instance, with the upcoming James Webb Space Telescope \citep{gardner2006james}.

The NASA \emph{Kepler} Mission \citep{borucki2010kepler} has discovered a large number of planets of intermediate size, with radii between Earth and Neptune, also known as sub-Neptunes. Since the size of planets is directly dependent on physical mechanisms in formation and evolution, the absence of sub-Neptunes in our solar system and their abundance among the exoplanet population has raised numerous fundamental questions. Many theoretical and statistical studies have been made on such planets. \cite{rogers2015most} showed that most planets with R > 1.6 R$_{\mathrm{E}}$ have low density and are inconsistent with a pure rocky composition. \cite{fulton2017california} demonstrated that the sub-Neptunes' distribution of radii is bimodal with 2 peaks centered at 1.2 R$_{\mathrm{E}}$ and 2.4 R$_{\mathrm{E}}$, which reveals a gap in planet radii between 1.5–2.0 R$_{\mathrm{E}}$. While this bi-modal distribution can be explained by photoevaporation  \citep{owen2013kepler,owen2017evaporation, lopez2014understanding} and core‐powered mass loss \citep{ginzburg2016super, ginzburg2018core} the composition and origin of close‐in sub‐Neptunes are not yet clear. Additional detections and precise characterizations are the keys to progress in answering our questions about their nature. One of the primary goals of TESS is to measure the mass of at least 50 transiting planets with a radius smaller than 4 Earth radii \citep{ricker2015jatis}. As of 22 April 2021, TESS has found more than 750 candidates \footnote{\url{https://tess.mit.edu/publications/}} (so-called TESS object of interest or TOI) with radii smaller than 4 R$_{\mathrm{E}}$. To date, 122 of them have been confirmed and have a mass measurement \citep[e.g.,][]{dragomir2019tess,gunther2019super,nielsen2020mass}.

Here, we announce the detection and characterization of a sub-Neptune orbiting a bright (V=8.4) K0 star using TESS photometric data and SOPHIE, APF, and HIRES RVs. In Sect. 2, we describe the variety of observations used to characterize the HD207897 (TOI-1611) system, including photometric, spectroscopic, and high-resolution imaging data. In Sect. 3, we analyze the data and present the characterization of the host star and planet, combining models on RVs and transit data. Finally, we present our discussion and conclude in Sect. 4.
%--------------------------------------------------------------------
\section{Observations}
\subsection{Photometry}
\subsubsection{TESS}
TESS planned to observe 80\% of the sky in 26 sectors for 2 years. Each sector is approximately 27 days, with partial overlap between sectors \footnote{\url{https://tess.mit.edu/observations}}. Because HD207897 is located near the north ecliptic pole where sectors overlap, it appears in many sectors.  As reported in the Web TESS Viewing
Tool (WTV)\footnote{\url{https://heasarc.gsfc.nasa.gov/cgi-bin/tess/webtess/wtv.py}}, HD207897 observations were taken on five sectors divided into two continuous periods from sectors 18-20 and 25-26, with a total time span of 131 days. After observations of sector 18 were completed, the MIT's Quick Look pipeline \citep[QLP-][]{2020RNAAS...4..204H,2020RNAAS...4..206H} detected the signature of two transits of HD207897 b at a period of 16.20 d, and an alert was issued 19 December 2019 by the TESS Science Office. No transits occurred during sector 19 as the sole transit fell in the data gap between the two orbits. After observations of sector 20 were downlinked, the Science Processing Operations Center \citep[SPOC;][]{jenkins2016tess} at NASA Ames Research Center conducted a transit search \citep{2002ApJ...575..493J,2010SPIE.7740E..0DJ} and detected two transits of HD207897 b at a signal-to-noise ratio (S/N) of 23.3. A limb-darkened transit model was fitted to the transit signature \citep{Li:DVmodelFit2019}, which passed all the diagnostic tests presented in the Data Validation report \citep{Twicken:DVdiagnostics2018}, including the odd-even transit depth test, the ghost diagnostic test, and the difference image centroiding test, which located the source of the transit signatures within 0.42$\pm$2.5 arcsec of the target star HD207897. Two additional transits were observed and detected in sector 25, and one more occurred during sector 26. A multi-sector search of sectors 18-26 by the SPOC detected 7 transit events of HD207897 b in total, at an S/N of 34.3 and an average depth of 913.7$\pm$21.1 ppm. No additional transiting planet signatures were detected in any of the SPOC or QLP runs.

We used the short-cadence (2 minutes) observations of TESS data from these 5 sectors in our photometric analysis, presented in Sect \ref{subsec:photmetryAnalysis}. These data were reduced by the SPOC Pipeline and are publicly available on Mikulski Archive for Space Telescopes (MAST) \footnote{\url{https://mast.stsci.edu/portal/Mashup/Clients/Mast/Portal.html}}.
 
\subsection{High-resolution spectroscopy}
\subsubsection{SOPHIE}

HD207897 had been monitored, before TESS observations, between 2012 to 2015 by the SOPHIE spectrograph \citep{perruchot2008sophie,bouchy2013sophie+}. The star measurements were conducted as a part of the Recherche de Planètes Extrasolaires (RPE) sub-program dedicated to detecting Neptunes and Super-Earths orbiting nearby and bright solar-type stars \citep{bouchy2009sophie,courcol2015sophie, hara2020sophie}. The observations were performed using high-resolution mode (resolution power of $\lambda / \Delta \lambda  \approx 75 000$) with a simultaneous Thorium-Argon (Th-Ar) calibration lamp, allowing us to monitor the instrumental drift. We collected 44 high-resolution spectra with a RV root mean square (RMS) of 6.9  m s$^{-1}$ (see Sect.~\ref{RV data reduction} for more details about the derivation of the RVs). A first analysis of the data in 2015 showed a periodicity close to 16.3 d. However, our preliminary analysis based on an erroneous estimate of the rotational period of the star led us to attribute the 16.3 d signal to stellar spot modulation at P$_{\mathrm{rot}}$ /2, and observations of the star were stopped. After the release of the TESS data, we resumed SOPHIE observations to gather more data to better separate the activity signal and planetary RV variation, and also investigate for other possible planets. The star was observed again in 2020 with a simultaneous Fabry-Perot (FP) reference spectrum, gathering 24 additional high-resolution spectra with RV RMS of 4.43 m s$^{-1}$. 

 Our final SOPHIE dataset includes 68 spectra. With an exposure time ranging from  1000 to 1500 s, we achieved a median signal-to-noise ratio (S/N) of 97.7 per pixel at 550 nm. The mean RVs uncertainty, computed using the quadratic sum of photon noise and wavelength calibration error, is 1.7 m s$^{-1}$. The RVs are presented in Table \ref{tab:rvs sophie}, after applying the corrections mentioned in Sect. \ref{RV data reduction}.

\subsubsection{HIRES}

From the summit of Maunakea, we first observed HD207897 using the Keck I telescope and HIRES spectrometer (Vogt 1994) from 2003 July 7 as part of the California Planet Search, and additional RVs were collected for eleven months beginning on 2020 Jan 21. Thirty-seven RVs were collected using the B5 decker (0.87" x 5.0") resulting in a resolution of 50,000. The median exposure time was 231 s, an average signal-to-noise ratio per pixel of 220, and internal uncertainty of 1.04 m/s. The Doppler pipeline and observing setup follows the California Planet Search procedure outlined in \cite{howard2010california}. The RV data set are presented in Table \ref{rvs hires}. We note that data before 2004 July 9 (3 data points) are used a different CCD detector. For sake of simplicity, we only used them in our RVs periodogram by applying an offset term (Fig.\ref{fig:all}). 

\subsubsection{APF}

From the summit of Mt. Hamilton at Lick Observatory, we collected 23 RVs of HD207897 using the Levy Spectrograph \citep{burt2014achieving} on the APF from 2 Jun 2020 until 28 Feb 2021. Twenty-three RVs were collected having a median exposure time of 1200s, a signal to noise per pixel of 86, and internal uncertainty of 3.0 m/s. With resolution of 100,000, this slit-fed spectrograph uses the iodine-cell technique to calculate RVs descended from \cite{butler1996attaining}. For additional details on the instrumental setup and data reduction see \cite{fulton2015three}. The full list of corrected RVs can be found in Table \ref{tab:rvs apf}.

\subsubsection{FIES}
From the Roque de los Muchachos Observatory in La Palma, Spain, we observed with the FIbre-fed \'{E}chelle Spectrograph \citep[FIES;][]{Telting2014} at the 2.56\,m Nordic Optical Telescope and obtained a total of 4 spectra between 27 December 2019 and 15 January 2020. We used the high-resolution fibre ($R \sim 67,000$) and extracted the spectra following \citet{Buchhave2010}. The S/N per resolution element at 550 nm ranges between 62 and 127. We used the FIES data for stellar classification (see section~\ref{subsec:spectralanalysis}) and decided not to include them in the RV analysis, since all the observations were acquired near phases 0.25 and 0.75.
 
\subsubsection{TRES}
From the Fred Lawrence Whipple Observatory (FLWO) atop Mt. Hopkins, Arizona, USA, we observed with the 1.5\,m Tillinghast Reflector Telescope using the Tillinghast Reflector Echelle Spectrograph (TRES; \citep{gaborthesis, TRES}) to obtain 1 spectrum (SNRe $\sim 36$) on UT 31 December 2019. TRES is a fiber-fed optical spectrograph with a resolving power of $R \sim 44,000$. The spectrum was extracted following the procedures outlined in \citet{Buchhave2010} and used to derive stellar parameters as described in 
section~\ref{subsec:spectralanalysis}. 
 
 \subsection{High-spatial resolution imaging}
 \label{sec:hr}
\begin{figure}
    \centering
    \includegraphics[scale=0.5]{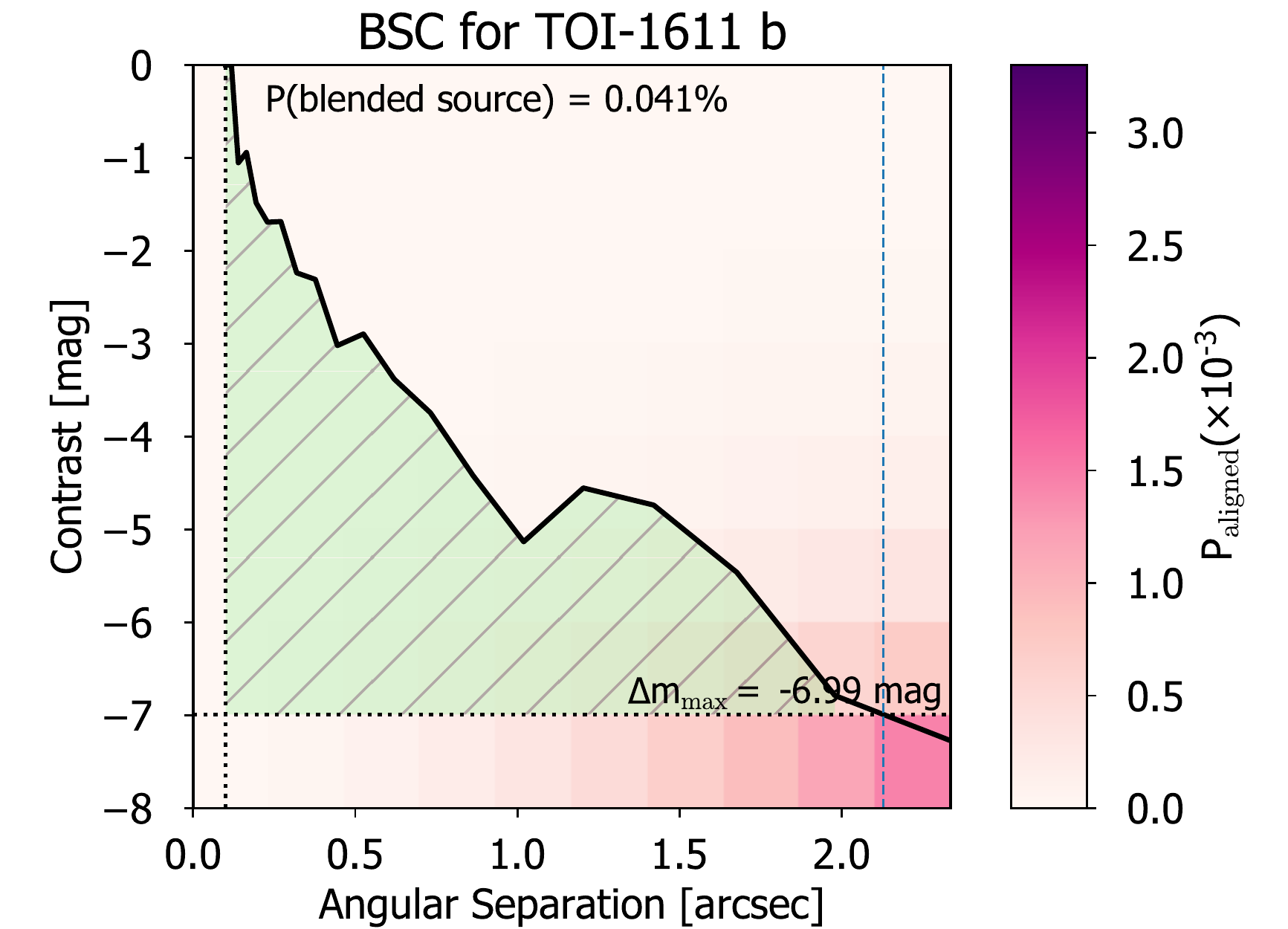} 
    \caption{Blended Source Confidence (BSC) curve from the AstraLux SDSSz image (solid black line). The color on each angular separation and contrast bin represents the probability of a chance-aligned source with these properties at the location of the target, based on TRILEGAL model (see Sect.~\ref{sec:hr} within the main text). The maximum contrast of a blended binary capable of mimicking the planet transit depth is shown as a dotted horizontal line. The green-shaded region represents the non-explored regime by the high-spatial resolution image. The BSC corresponds to the integration of $P_{\rm aligned}$ over this shaded region.}
    \label{fig:bsc}
\end{figure}

We used the AstraLux high-spatial resolution camera \citep{hormuth08} at the 2.2\,m telescope of the Calar Alto Observatory (CAHA, Almeria, Spain) to unveil possible close-in companions to HD207897. This instrument applies the lucky imaging technique to retrieve diffraction limited images from point source objects by acquiring thousands of short-exposure frames that freeze atmospheric variations and thus produce diffraction limited images. We observed this target on the 26th of February 2020 under good weather conditions and a mean seeing of 0.9 arcsec. Given the brightness of this target, we obtained 11\,000 frames with an individual exposure time of 10 milliseconds. We used the Sloan Digital Sky Survey z filter (SDSSz), which is the optimal one to obtain the highest possible resolution with AstraLux. We also used a field-of-view windowed to $6\times6$ arcsec. All frames were reduced by the instrument pipeline  \citep{hormuth08}, which performs the basic reduction (bias subtraction and flat field correction), aligns all frames,  and measures the Strehl ratio \citep{strehl1902} of the individual images. This metric is then used to select the best images to perform the final stacking. We used the 10\% best frames to produce the final high-resolution image. The final image does not show signs of close companions. We performed a dedicated search by removing the instrumental point spread function scaled to the target peak flux and no additional sources were present in the image. We then followed the procedures described in \cite{lillo-box12,lillo-box14b} by using our own developed \emph{astrasens} package\footnote{\url{https://github.com/jlillo/astrasens}} to obtain the sensitivity of our image by performing an injection/retrieval of artificial sources. The sensitivity curve is shown in Fig \ref{fig:bsc}. 

\begin{figure}
    \centering
    \includegraphics[scale=0.5]{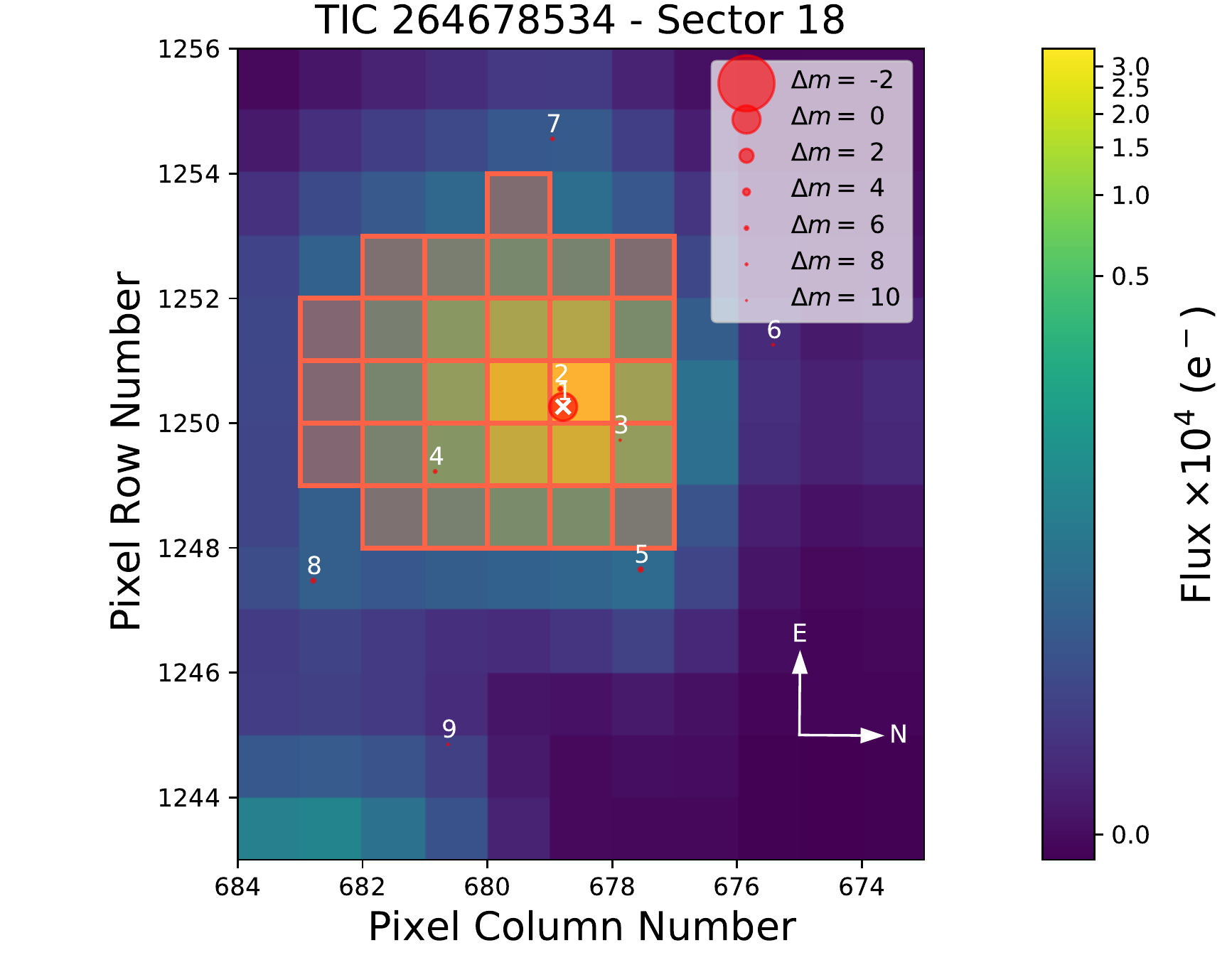} 
    \caption{Target Pixel File of TESS for HD207897 corresponding to sector 18, created with \texttt{tpfplotter} \citep{aller2020planetary}. The red square is related to the SPOC aperture mask and red circles indicate Gaia DR2 magnitude of nearby stars with the size of depending on their brightness. HD207897, number 1, has a Gaia(mag) = 8.1 while the nearest star TIC 264678535, number 2, has a Gaia(mag)=13.8. }
    \label{tpf}
\end{figure}

By using this sensitivity curve, we additionally estimated the probability of an undetected blended source in our high-spatial resolution image (BSC, see procedure described in \citealt{lillo-box14b}). We use a python implementation of this approach (\texttt{bsc}, by J. Lillo-Box) which uses the TRILEGAL\footnote{\url{http://stev.oapd.inaf.it/cgi-bin/trilegal}} galactic model \citep[v1.6][]{girardi12} to retrieve a simulated source population of the region around the corresponding target\footnote{This is done in python by using the \emph{astrobase} implementation by \cite{astrobase}.}. This simulated population is used to compute the density of stars around the target position (radius $r=1^{\circ}$) and derive the probability of chance-alignment at a given contrast magnitude and separation. In this case, given the AstraLux high-resolution image, we obtain a probability of an undetected blended source of 0.041\%. We consider this probability low and the odds that such an undetected source being an appropriate binary capable of mimicking the transit signal to be negligible. In addition, by using \texttt{tpfplotter} code\footnote{\url{https://github.com/jlillo/tpfplotter}} \citep{aller2020planetary}, we plotted the Target Pixel File from TESS for HD207897 b (see Fig.\ref{tpf}). We checked for possible light contamination by considering the nearby stars TIC 264678535 ($T_{mag}=13.3$), TIC 264678538 ($T_{mag}=15.69$) and TIC 264678529 ($T_{mag}=14.39$), all inside the selected apertures analyzed by PDC-SAP. Since the Gaia G band-pass is quite similar to TESS band-pass, we used the Gaia fluxes of these stars and estimated the level of contamination. The total flux due to nearby stars was only $0.8\%$ of HD207897's flux which is automatically corrected by SPOC.
Given the low probability of undetected blended sources from the AstraLux high-resolution image, negligible amount of light contamination due to nearby stars, and similar Keplerian amplitudes derived with different masks (see Sect. \ref{RV data reduction}), we conclude that the transit and RV variations originate from the main target within the TESS TPF (HD207897), and are of planetary origin.

\section{Analysis and Results}
\label{sec:analysisAndResults}
\subsection{Stellar parameters}
\label{subsec:spectralanalysis}

 To obtain the stellar atmospheric parameters, we first summed the 66 spectra of SOPHIE after correcting from RV variations of the star, barycentric Earth radial velocity, and background correction of calibration lamps \citep{Hobson2019}. It resulted in a high S/N per pixel spectrum of 772 at 550 nm.

 The $T_{eff}$ and [Fe/H] were calculated following the procedure described in \cite{santos2013sweet} and \cite{sousa2018sweet}. It is based on the equivalent width of Fe I and Fe II lines and assumes a balance in excitation and ionization of iron lines in local thermal equilibrium. Abundances of Mg and Si were derived closely following the curve-of-growth analysis methods described in our previous works \citep{Adibekyan-12, Adibekyan-15}. Abundances of C and O are very difficult to determine for stars cooler than about 5200 K \citep{Delgado-10, Bertrandelis-15}. We estimated the abundances of these elements empirically by using a machine learning algorithm (we used the estimator "RandomForestRegressor") from the Python Scikit-learn package \citep{Pedregosa-11}. The estimation of C and O was based on the abundance of Mg and Fe \citep{Delgado-17}. Our initial sample was based on the HARPS sample for which abundances of Mg and Fe are available \citep{Adibekyan-12}. Then we derived O abundance for 535 stars and C abundance for 758 stars following the methodology of our previous works \citep{Delgado-10, Bertrandelis-15}. These samples were used as training and test datasets. The resulting abundances and other stellar parameters of HD207897 are presented in Table \ref{stellar parameters}.
 
 As an independent Stellar Classification analysis, we used the FIES and TRES data following \citep[SPC;][]{Buchhave2012,Buchhave2014}. The FIES analysis uses five spectral orders spanning a wavelength range from 5065 to 5320 \AA{}, while the TRES analysis uses three spectral orders spanning the range from 5060 to 5300 \AA{}. The spectra were compared to a library of synthetic templates, to measure the effective temperature $T_{\rm eff}$, surface gravity $\log g$, projected rotational velocity $v \sin i$, and metallicity [m/H] (a solar mix of metals). We analyzed the spectra individually and calculated a weighted average of each parameter. The results were $T_{\rm eff}=5085\pm 50{\rm K}$, $\log g = 4.48\pm 0.10$, $v \sin i < 2 {\rm km s}^{-1}$, and [m/H]$= -0.23\pm 0.08$ which are in agreement with the other methods.

 \begin{figure}
 	\includegraphics[width=\columnwidth]{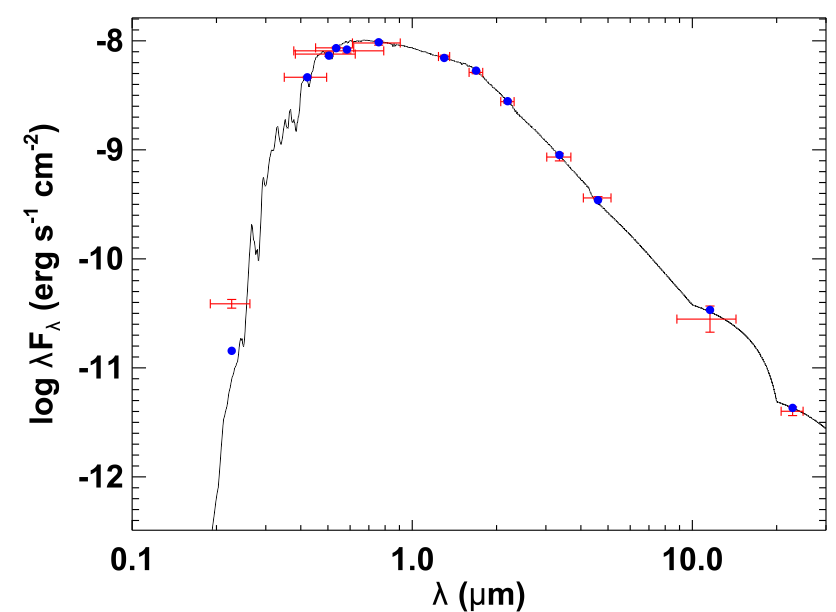}
	\caption{Spectral energy distribution of HD207897. Red symbols represent the observed photometric measurements, where the horizontal bars represent the effective width of the passband. Blue symbols are the model fluxes from the best-fit Kurucz atmosphere model (black).}
	\label{fig:SED} 
	\end{figure}

 %\subsection{Stellar parameters}
 %\label{stellar_paramj}
 We also performed an analysis of the broadband spectral energy distribution (SED) of the star together with the {\it Gaia\/} EDR3 parallax \citep[with no systematic offset applied; see, e.g.,][]{stassun2021parallax}, in order to determine an empirical measurement of the stellar radius, following the procedures described in \citet{stassun2016eclipsing,stassun2017accurate,stassun2017empirical}. We pulled the $B_T V_T$ magnitudes from {\it Tycho-2}, the $JHK_S$ magnitudes from {\it 2MASS}, the W1--W4 magnitudes from {\it WISE}, the $G G_{\rm BP} G_{\rm RP}$ magnitudes from {\it Gaia}, and the NUV magnitude from {\it GALEX}. Together, the available photometry spans the full stellar SED over the wavelength range 0.2-22~$\mu$m (see Figure~\ref{fig:SED}).  

 We performed a fit using Kurucz stellar atmosphere models, with the effective temperature ($T_{\rm eff}$), metallicity ([Fe/H]), and surface gravity ($\log g$) adopted from the spectroscopic analysis. We also fixed the extinction $A_V \equiv 0$ based on the star's proximity (see Table \ref{stellar parameters}). The resulting fit (Figure~\ref{fig:SED}) has a reduced $\chi^2$ of 1.1, excluding the {\it GALEX} NUV flux which indicates a moderate level of activity. Integrating the (unreddened) model SED gives the bolometric flux at Earth, $F_{\rm bol} = 1.401 \pm 0.016 \times 10^{-8}$ erg~s$^{-1}$~cm$^{-2}$. Taking the $F_{\rm bol}$ and $T_{\rm eff}$ together with the {\it Gaia\/} parallax, gives the stellar radius, $R_\star = 0.785 \pm 0.014$~R$_\odot$. In addition, we can estimate the stellar mass from the empirical relations of \cite{torres2010accurate}, giving $M_\star = 0.84 \pm 0.05$~M$_\odot$. 
  Additionally, we cross-checked these values with a different methodology by EXOFASTv2 in Sect \ref{exofast}.

\begin{table}
\caption{\label{stellar parameters} Stellar properties of HD207897}
\centering
\resizebox{\columnwidth}{!}{%
\begin{tabular}{lll}
\hline
\hline
Other identifiers &  & \\
 & TIC 264678534 & \\
 & HD 207897 & \\
 & BD +83 617 & \\
 & HIP 107038 & \\
 & Gaia DR2 2300641567596591488  & \\
 & 2MASS J21404490+8420005    &  \\
 \hline
\hline
Parameter & HD207897 & References\\
\hline

 & Astrometric properties &\\
  & &\\
Parallax (mas) & $  35.3446 \pm  0.0468 $ & $ Gaia DR2 ^{*}$\\
& $35.3581  \pm  0.0159 $ & $ Gaia EDR3$\\
Distance & 28.25 $\pm$ 0.03 & Gaia DR2\\
$  \alpha (h m s) $ & $ 21: 40: 44.78 $ & Gaia DR2\\ 
$ \delta(d m s) $ & $ +84: 20: 00.56 $ & Gaia DR2 \\ 
  & &\\
 & Photometric properties &\\
   & &\\
 B-V & $0.86\pm0.02$ &  HIP\\
 V(mag) & $8.37\pm 0.0015$  & HIP \\
 Gaia(mag) & $ 8.1304\pm 0.0004$ & Gaia DR2 \\
$ Gaia_{BP} (mag)$&  $8.6051\pm 0.0025$& Gaia DR2 \\
$Gaia_{RP} (mag)$& $7.5402\pm 0.0014$& Gaia DR2 \\ 
TESS(mag) & $7.58 \pm 0.006 $ & TESS \\
 J(mag) &  $ 6.830\pm 0.023 $ & 2MASS \\
 H(mag) & $ 6.391 \pm 0.034 $ & 2MASS \\
 $K_{s}(mag)$ & $ 6.312 \pm 0.026 $ & 2MASS \\
$W_{1}(mag)$ & $ 6.262 \pm 0.088 $ & WISE \\
$W_{2}(mag)$ & $ 6.219 \pm 0.025 $ & WISE \\
$ W_{3}(mag)$ & $6.275\pm 0.015$ & WISE \\
$ W_{4}(mag)$ & $6.233 \pm 0.045$ & WISE \\
  & &\\
 & Spectroscopic properties &\\
   & &\\
 Spectral type & K0V &  HIP\\
   $\xi_{t}$   $ (km s^{-1})$  &$0.53 \pm  0.10$& Sec \ref{subsec:spectralanalysis}\\
 $\log (R'_{HK})$ & $-4.83\pm0.10 $ & Sec \ref{Stellar rotation and activity}\\
 $v\sin i $ $ (km s^{-1})$& $< 2$& Sec \ref{Stellar rotation and activity}\\
 $[Fe/H] $ $dex$  &$-0.21 \pm  0.02$& Sec \ref{subsec:spectralanalysis}\\
 $[C/H] $ $dex$  & $-0.23\pm 0.07 $&  Sec \ref{Stellar rotation and activity}\\
$[O/H] $ $dex$  & $-0.11 \pm 0.08 $&  Sec \ref{Stellar rotation and activity}\\
$[Mg/H] $ $dex$  & $-0.17\pm 0.06 $&  Sec \ref{Stellar rotation and activity}\\
$[Si/H] $ $dex$  & $-0.19\pm 0.05 $&  Sec \ref{Stellar rotation and activity}\\

 & &\\ 
  & Bulk properties &\\
  & &\\
 Mass ($ M _{sun} $)& $0.80^{+0.036}_{-0.030}$ & Sec \ref{exofast}   \\
 &$0.84\pm 0.05$& Sec \ref{subsec:spectralanalysis}\\
 Radius($ R _{sun} $) & $0.779^{0.019}_{-0.018}$ & Sec \ref{exofast} \\
 & $0.785 \pm 0.014$& Sec \ref{subsec:spectralanalysis}\\
 $\log{g} (cgs)$ & $4.559^{+0.026}_{-0.025}$& Sec \ref{exofast} \\
 $ L_{s}(L_{sun}) $ & $0.360^{+0.019}_{-0.014}$& Sec \ref{exofast}\\
 $T_{eff} (K)$ &$5070^{+60}_{-57}$&Sec \ref{exofast} \\
 $ P_{rot}(days) $ & $ 37 \pm 7$ & Sec \ref{Stellar rotation and activity}\\
\hline
\end{tabular}
}
\tablefoot{$^{*}$ We applied the offset correction as prescribed in \cite{lindegren2018gaia}.}

\end{table}

\subsection{Photometry data analysis}
\label{subsec:photmetryAnalysis}

\begin{figure*}[h!]
	\centering 
	\includegraphics[width=0.9\textwidth]{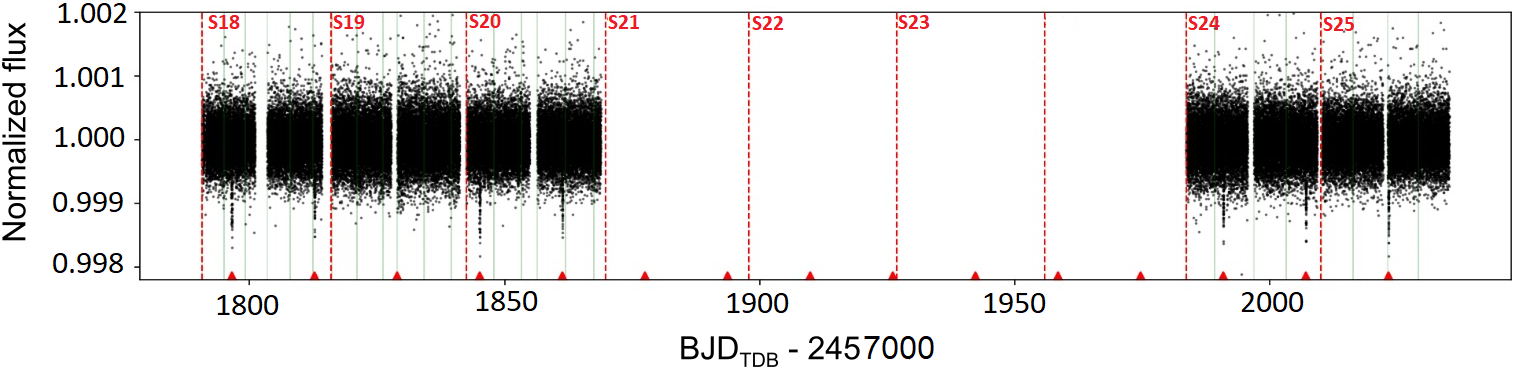}
   	\begin{tabular}{cc}
	  \includegraphics[width=.45\textwidth]{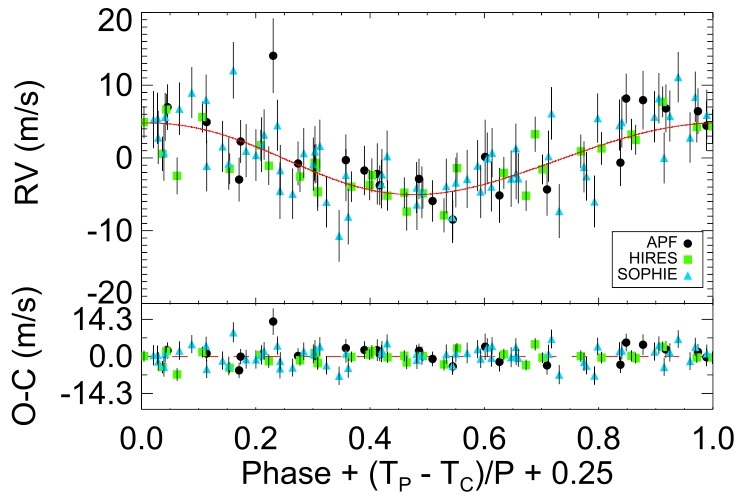} 
	  \includegraphics[width=.45\textwidth]{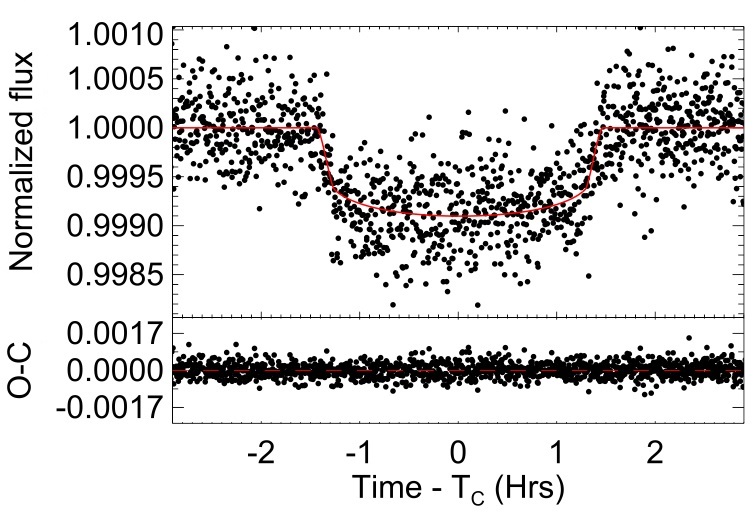} 
	  \end{tabular}
	\caption{TESS light curves and EXOFASTv2 best fit models on RVs and phometric data. \emph{Top panel}: The Full PDC-SAP (2-minute) TESS light curve after detrending, taken from sectors 18, 19, 20, 25 and 26. The red triangles indicate the transit events, and the green vertical lines indicate the momentum dumps of spacecraft which occurs every 2.5 days. None of the transit events fall during momentum dumps. \emph{Bottom left}: Phase-folded SOPHIE, APF, and HIRES RVs of HD207897 b at the period of 16.20 d. \emph{Bottom right}: TESS Phase-folded light curve.}
	\label{fig:lightcurve} 
\end{figure*}

 The photometry was extracted with the Pre-search Data Conditioned Simple Aperture Photometry (PDC-SAP) pipeline \citep{Stumpe2012,smith2012kepler, stumpe2014multiscale} provided by the TESS team. We removed nans, flagged low-quality data, and $5 \sigma$ outliers. We then normalized and detrended the light curve with a spline robust iterative sigma-clipping method \citep{schoenberg1946contributions} using the Wotan package \citep{hippke2019wotan} \footnote{\url{https://github.com/hippke/wotan}}. This method detrends the light curve by fitting the spline through minimizing the sum of squared residuals along with iteratively sigma-clipping. This step is one of the most important steps in the photometric analysis as it reduces the number of false-positive signals by removing instrumental and stellar noise. On the other hand, with such detrending algorithms, there is always a risk of changing the transit depth or even fully removing shallow transits. For this reason, we tested different values of knots and chose the value of 0.7, which was the maximum value that seemed enough for removing the light curve variabilities. The resulting light curve is shown in Figure \ref{fig:lightcurve} with the transit events marked with red triangles.

To look for any periodic signals in the data, we used the transit-least-squares (TLS) algorithm \citep{hippke2019optimized}, a method to investigate planetary transits taking into account the stellar limb-darkening \citep{mandel2002analytic} and also the effects of planetary ingress and egress. This method is publicly available \footnote{\url{http://github.com/hippke/tls}} and has been optimized for Signal Detection Efficiency (SDE) of small planets. We searched for periodic signals in the range 0.6 to 122 d and sampled 28607 periods within that range. The result showed a prominent periodic signal occurring every 16.20 d with a S/N of 43, giving a SDE of 76.2 and a false alarm probability (FAP) below 0.01 \% (see Fig \ref{fig:tls-sed}). After masking the signal at 16.20 d, we ran the TLS again and did not see any periodic signals. A search for a long-period beyond 122 d did not show any significant signals either.

\subsection{Radial velocity data analysis}
\label{radial_velocity analyze}

\subsubsection{RVs data reduction on SOPHIE}
\label{RV data reduction}

The SOPHIE data were reduced with the SOPHIE Data Reduction pipeline \citep[DRS,][]{bouchy2009}, which extracts the RV by cross-correlating spectrum with a binary mask and then doing a Gaussian fit of the cross-correlation function (CCF) \citep{pepe2002coralie}. We tested different masks, including G2, K0, and K5. Since they showed similar Keplerian amplitude variations, it is unlikely that these variations are produced by blend scenarios composed of stars of different spectral types \citep{bouchy2008radial}. We finally adopted the RV data derived with the K5 mask as it presented a smaller RV dispersion. 

We subsequently excluded 6 RV points that did not reach the required quality: 3 points with a lower signal-to-noise than required $ S/N_{550} > 50$, 2 points with moonlight pollution, and 1 point both being an outlier and resulting from an observation performed with no simultaneous calibration. We corrected the remaining points for the charge transfer inefficiency (CTI) effect \citep{santerne2012sophie}. This correction ranged from 1.4 to 6 m s$^{-1}$ with a mean RV correction of 2.1 m s$^{-1}$. Afterward, we removed the nightly drift of the telescope measured from the simultaneous Th-Ar or FP reference spectra. This correction also falls within the 0.1 to 9 m s$^{-1}$range, with a mean RV of 2.1 m s$^{-1}$.  

The next step was removing a long-term drift of the zero point due to the instrumental effect identified in SOPHIE RV data \citep[][]{courcol2015sophie,hobson2018sophie}. To track this offset, so-called 'constant' stars have been monitored each night. We combined these observations to build a time series of the RV master constant (see \cite{courcol2015sophie} for more details on the method). We then estimated long-term zero point drift as a function of time for each night of observation and subtracted it from the HD207897 RV data. This correction is in the 0.2 to 6 m s$^{-1}$ range with a mean value of 3.3 m s$^{-1}$. Since we saw the effects of the long-term drift of the zero points on the bisector span of constant stars, we applied the same processes to the bisector of HD207897 as well. The correction for bisector was between 4 to 11 m s$^{-1}$ with a mean value of 2 m s$^{-1}$. Once these corrections were applied, the HD207897 RVs reduced from an original RMS of 6.2 m s$^{-1}$ to a final RMS of 4.9 m s$^{-1}$.
\begin{figure}
%	\centering 
	\includegraphics[width=\columnwidth]{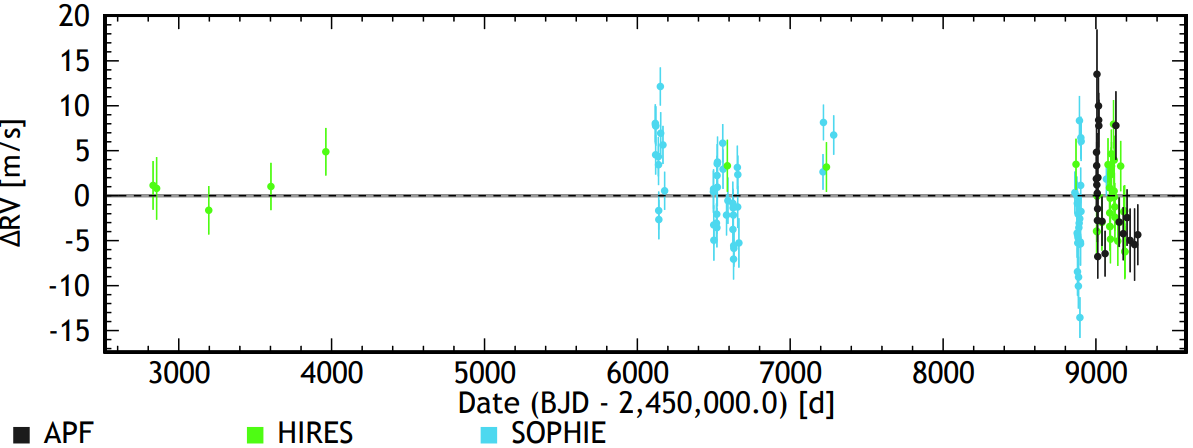}
	\caption{Radial velocities of HD207897 from SOPHIE, APF and HIRES, before linear trend correction.}
	\label{fig:Radial velicity} 
	
\end{figure}
\begin{figure}[h!]
    \begin{tabular}{@{}c@{}}
 		\includegraphics[raise=-\dp\strutbox,width=0.47\textwidth]{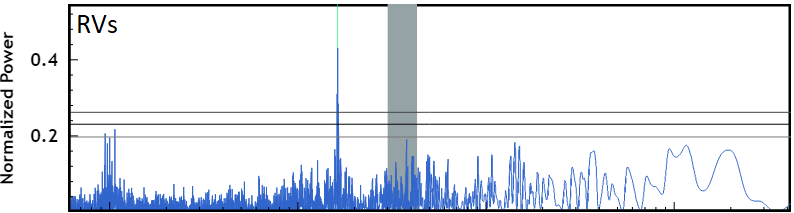}\\
 		\includegraphics[width=0.47\textwidth,raise=-\dp\strutbox]{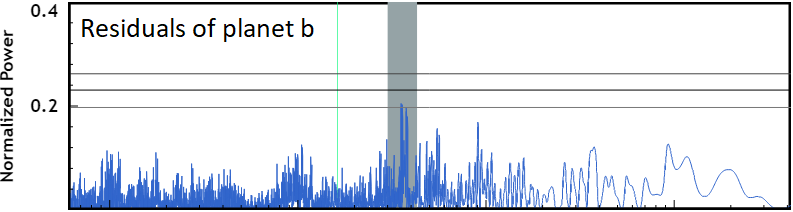}\\
 		\includegraphics[width=0.47\textwidth,raise=-\dp\strutbox]{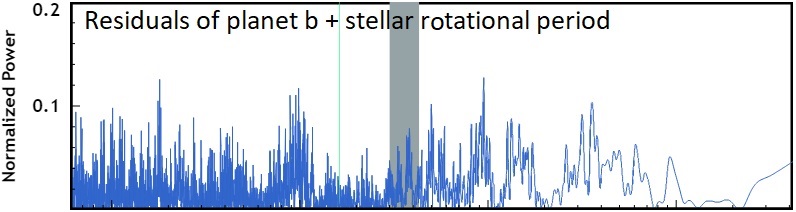}\\
 		\includegraphics[width=0.47\textwidth,raise=-\dp\strutbox]{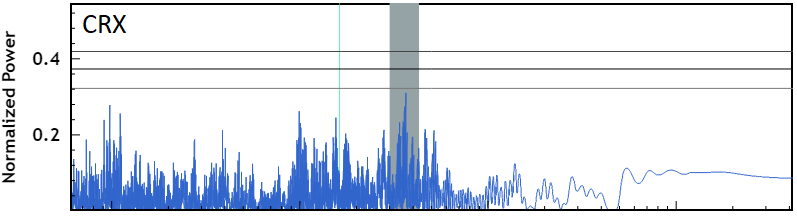}\\
 		\includegraphics[width=0.47\textwidth,raise=-\dp\strutbox]{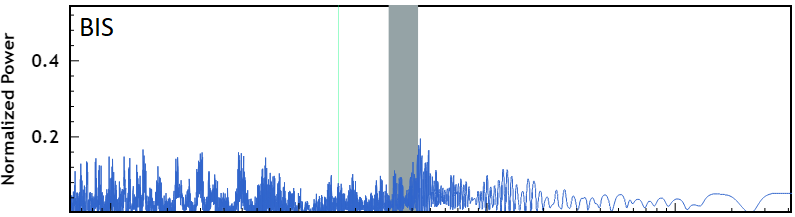}\\
 		\includegraphics[width=0.47\textwidth,raise=-\dp\strutbox]{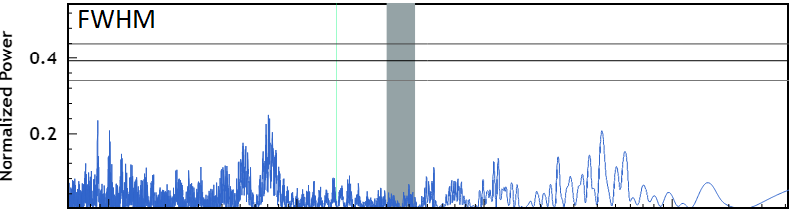}\\
 		\includegraphics[width=0.47\textwidth,raise=-\dp\strutbox]{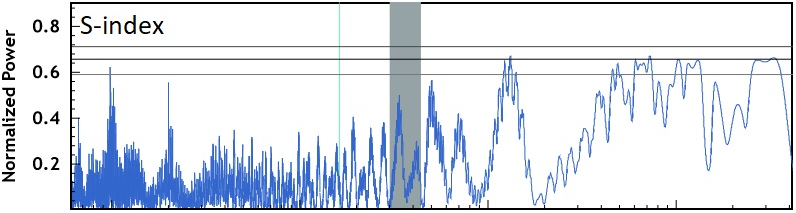}\\
 		\includegraphics[width=0.47\textwidth,raise=-\dp\strutbox]{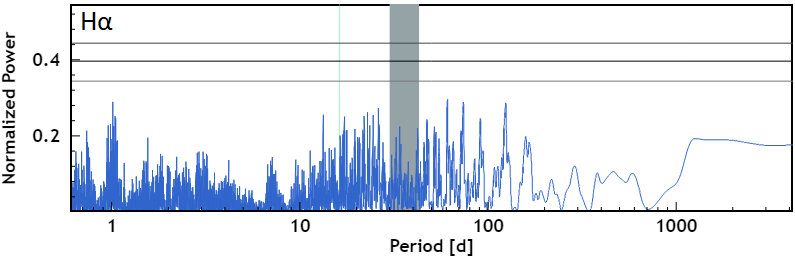}
 	\end{tabular}
	\caption{ Normalized periodograms of RVs and activity indexes for HD207897. From \emph{top to bottom panel}: RVs, residual of RVs after Keplerian fit on 16.20 d, Residuals of fit on planet and stellar rotation signal, CRX, bisector, FWHM, and $H \alpha$ index. The cyan vertical line marks the position of the highest peak at RVs periodogram at 16.20 d and shows no corresponding peak in the stellar activity periodograms. Horizontal lines indicate 0.1, 1, 10 $\%$ FAP level, from top to bottom, respectively. Also, the grey vertical strip is highlighting the position of the estimated rotational period of star in Sect. \ref{Stellar rotation and activity}}
	\label{fig:all}
\end{figure}
\subsubsection{Stellar rotation and activity}
\label{Stellar rotation and activity}

To investigate the activity of the star, we used several indicators such as bisector span, CCF FWHM, Chromatic RV index (CRX), $H\alpha$, and $\log (R'_{HK})$ from SOPHIE spectra as well as S-index from the HIRES spectra.

We obtained the bisector span and CCF FWHM from the SOPHIE data reduction system. The CRX is extracted using SERVAL \citep{zechmeister2018spectrum} code. To compute the $H\alpha$ index, which measures the flux in the $H\alpha$ line, we followed the definitions of \cite{boisse2011disentangling}. To do so, we also applied background correction to the data as appropriate for the two different calibration lamps \citep[for further details, see][]{Hobson2019}. 

We derived the S-index from HIRES spectra following \cite{baliunas1995chromospheric} and \cite{paulson2002searching}. We also computed the $\log (R'_{HK})$ index on SOPHIE spectra, following the method of \cite{noyes1984rotation}, but the S/N is low in all first (bluest) orders, where Ca H\&K lines are located, with one exception in 2020 (S/N =33). We decided to sum the bluest order of spectra with S/N > 20 to reach to better S/N ratios and then calculate $\log (R'_{HK})$. We first did so for the 2012-2015 spectra, of which 22 had a S/N > 20, giving a value of $\log (R'_{HK}) =-4.83\pm 0.10$. Only one 2020 spectrum presents a S/N > 20, so no summation was possible. This spectrum yields a value of $\log (R'_{HK})= -4.78 \pm 0.10$, indicating that the star may have a similar level of activity as during the 2012-2015 period. Our value of $\log (R'_{HK})$ for 2012-2015 is consistent with $\log (R'_{HK})= -4.86$ reported by \cite{brewer2016spectral}. This value indicates a modest activity level of HD207897.

To check our $\log (R'_{HK})$ value, we used the star's {\it GALEX} NUV flux excess (Fig.~\ref{fig:SED}). The observed NUV excess implies a chromospheric activity of $\log R'_{rm HK} = -4.82 \pm 0.05$ via the empirical relations of \cite{findeisen2011stellar}, consistent with the value obtained spectroscopically. Moreover, the NUV estimated activity implies an age of $\tau_\star = 4.4 \pm 0.9$ Gyr via the empirical relations of \cite{mamajek2008improved}.

We estimated a stellar rotation period of $ 37 \pm7$ d with the method proposed in \cite{noyes1984rotation}, which is in good agreement with an estimated period of $ 36 ^{+5}_{-4}$ d following \cite{mamajek2008improved}. These estimates are consistent with the value of 38 days from \cite{isaacson2010chromospheric}.

We sought to constrain the $P_{rot}$ of the star with RVs and activity indicators. We subtracted a linear drift in the CCF FWHM, CRX, and RVs time series, and a cubic drift in the S-index. Also, an offset term is fitted on $H\alpha$ data between the two different calibration lamps background correction (before and after BJD= 57284.4185). Then, we computed the periodogram of RVs, the RV-residuals of keplerian models, and activity indicators (Fig \ref{fig:all}). The RV-residuals of planet b show two peaks at 35.9 and 37.6 days near to 10 \% FAP.

We ran an $\ell_1$ periodogram on the RVs (see Fig. \ref{fig:l1-periodogram}) for comparison. This one is obtained with the same procedure as in \cite{hara2020sophie}: we considered alternative noise models characterised by an autocovariance function that is a sum of white, correlated, and quasi-periodic components with different amplitudes and time scales and ranked them with cross-validation. The correlated component is a Gaussian kernel characterised by its time-scale and amplitude, the quasi-periodic component is a Gaussian Kernel as in \cite{haywood2014} characterised by its decay time-scale, period, and amplitude. The amplitudes of the white, correlated, and quasi-periodic terms are taken on a grid (0 to 3 m/s with a 0.5 m/s step), the time-scales 0,3,6 days for the correlated term and 30, 60, 90 days for the quasi-periodic term and 37.6 days for the period of the quasi-periodic component. We also included in the base model one offset per instrument. Figure \ref{fig:l1-periodogram} corresponds to the highest ranked noise model. We found a significant signal at the planet period (FAP of $2\cdot 10^{-9}$) and found a signal at 37.6 d with a FAP of 0.5. The signal at 37.6 d very likely corresponds to a stellar signal.

While the periodogram of CCF FWHM and $H\alpha$ activity indicators do not exhibit any significant signals, CRX shows a peak near to 10 \% FAP at 36.5. The S-index periodogram shows several peaks at the long periods, however, no correlation (R= 0.16) is found between the S-index and HIRES RVs. We note that given the relatively small value of $v\sin i $ ($2\pm 1$ km s$^{-1}$), we could not see any correlations between the RVs and its residuals and bisector. 

We searched the SAP and PDC-SAP light curves for a signal of the rational period of the star in the photometry. For this, we applied the Systematics-insensitive Periodogram (SIP) method\footnote{\url{https://github.com/christinahedges/TESS-SIP }} \citep{angus2016systematics,hedges2020systematics} for SAP, and Gaussian process (GP) model on PDC-SAP light curves. The SIP method detrends SAP light curve (see Fig. \ref{fig:SIP} bottom) from TESS instrument systematics, along with calculating Lomb–Scargle periodogram and without requiring pre-detrending light curves as other methods \citep[e.g. AutoCorrelation Function (ACF) by][]{mcquillan2013measuring}. This method was initially used for the \emph{Kepler} mission and has recently been successfully applied for TESS data such as TOI-1259A \citep{martin2021toi} and TOI-700 \citep{hedges2020systematics}. The SIP periodogram (see Fig. \ref{fig:SIP} top) does not exhibit any significant signal for HD207897 b. Also applying a GP model on PDC-SAP light curves did not display any convincing signal either.

To examine further the origin of the 37.6 d signal, following \cite{hara2021b}, we check for the phase and amplitude consistency of the 37.6 days signal in the SOPHIE RVs. We use the statistic defined in Eq. 14 of  \cite{hara2021b}. We fit an offset and a linear trend as well as a sinusoidal model for the 16.2-day planet. Adopting the value of the RV jitter of 3.16 m s$^{-1}$, we find that the hypothesis that the phase and amplitude of the 37.6-day signal is constant is rejected at $2\sigma$, which further supports the hypothesis that this signal is due to activity.

Given the estimated stellar rotational period value and the signal at the CRX activity indicator, it is likely the signals at 35.9 and 37.6 days are due to the stellar rotation period. We take into account these peaks in our joint model analysis with EXOFASTv2 in Sect \ref{exofast}.

\subsubsection{Radial velocities results}
\label{RV results}

The combined RV data of HD207897 are plotted in Fig. \ref{fig:Radial velicity} after fitting a zero-points offset for each RV dataset (see table \ref{tab:hd207897.}). We removed a linear trend of $-0.28\pm0.11$ m s$^{-1}$ yr$^{-1}$ from RVs. We note that no notable differences were found in our results by including/excluding this linear drift. However, since we saw a clear linear drift in the activity indicators such as the CCF FWHM and CRX we decided to keep the drift.  

We investigated the footprints of the planet in RVs by searching for periodic signals. To do so we used the Data and Analysis Center for Exoplanets \citep[DACE,][]{delisle2016analytical} web platform\footnote{Available at https://dace.unige.ch}
and computed periodograms for the RVs (Fig: \ref{fig:all}). 

The RVs periodogram displays a clear peak at 16.20 d with a value below 0.1 \% false alarm probability \citep{baluev2008assessing}. Moreover, there is no corresponding peak in the periodogram of activity indicators, which shows that this periodic signal is likely due to a planet and not due to stellar activity, Figure \ref{fig:all}.

\begin{figure}
%	\centering
    \includegraphics[width=\columnwidth]{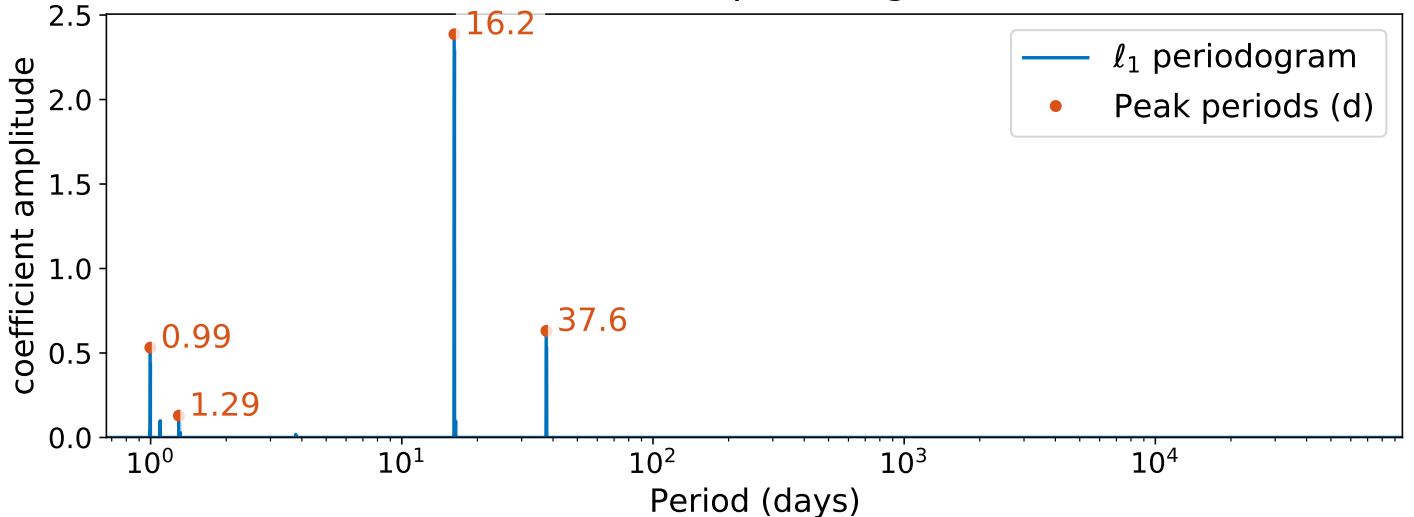}
	\caption{$\ell_1$ periodogram of SOPHIE RVs for HD207897 b, following \cite{hara2017radial}. The main peaks of the periodogram are indicated} with orange circles.
	\label{fig:l1-periodogram}
\end{figure}

 As it is shown in Figure \ref{fig:l1-periodogram}, the $\ell_1$ periodogram confirms that the most prominent signal appears at 16.20 days with FAP of $2\cdot 10^{-9}$. The value of the second-highest peak at 37.7 days is in agreement with the estimated rotational period of the star (see Section \ref{Stellar rotation and activity}). 

\subsection{Joint modeling of RV and photometry}
\label{exofast}

To explore all the parameters of the system, including both the host star and the planet, we modeled simultaneously and self-consistently the photometric observations of the star from five sectors of TESS, and the RV observations from SOPHIE, HIRES, and APF using the Fast Exoplanetary fitting package \citep[EXOFASTv2 \footnote{EXOFASTv2 is available at \url{https://github.com/jdeast/EXOFASTv2.}},][]{eastman2013exofast,eastman2017exofastv2,eastman2019exofastv2}. This global modeling software uses a differential evolution Markov Chain coupled with a Metropolis-Hastings Monte Carlo sampler that uses error scaling for exploring the system parameters.

EXOFASTv2 fits a total of 33 free parameters for the HD207897 system, which can be divided into the following categories:
\begin{itemize}
    \item twelve parameters related to the planet and activity period: the mid-transit time $T_{C}$, planet orbital period $P$, the planet-to-star radius ratio $R_{p}/R_{star}$ (only for planet), the orbital inclination $i$ (only for planet), the RV semi-amplitude $K$, and two more free parameters related to eccentricity $e$.
    \item Two parameters for each RV instrument: instrumental offset and jitters. One free parameter is also fitted for drift on RVs.
    \item Two limb darkening coefficients for TESS photometric bands pass, along with baseline flux and variance are fitted for the transit light curve.
    \item Eleven stellar parameters: Stellar mass $M_{*}$, Stellar radius $R_{*}$ and effective temperature $ T_{eff} $ by the MIST model, Stellar radius $R_{*,SED}$ and effective temperature $ T_{eff, SED} $ by the SED model, observed metallicity $[Fe/H]$, theoretical metallicity at the star's birth $[Fe/H]_{0}$ by MIST model, Age , equivalent evolutionary point $EPP$, V-band extinction $A_{v}$ and distance $d$.
\end{itemize}

Before running EXOFASTv2, we set a Gaussian prior on $ T_{eff} $ and [Fe/H] from our spectral analysis results, presented in Sect \ref{subsec:spectralanalysis}. We also imposed a prior on the Gaia DR2 parallax after applying the offset correction as described in \cite{lindegren2018gaia}. Broadband photometry presented in Table \ref{stellar parameters} is also included. We enforced an upper limit for the V-band extinction ($A_{v}$) from \cite{schlegel1998maps} and \cite{schlafly2011measuring}. The broadband photometry, the Gaia parallax, and the $A_{v}$ allowed us to model the stellar spectral energy distribution, a key to constrain the stellar radius. Indeed, EXOFASTv2 interpolates a pre-compiled 4d grid of bolometric corrections (logg, $ T_{eff} $, [Fe/H], reddening) to arrive at the broadband photometry flux directly. We did not set the limb darkening to let EXOFASTv2 constrain the best parameters of a quadratic limb darkening through the \cite{claret2017limb} tables for TESS bands and the stellar atmosphere's parameters ($ T_{eff} $, [Fe/H] index and log $g_{*}$). We used the Mesa Isochrones and Stellar Tracks evolutionary model \citep[MIST,][]{dotter2016mesa, choi2016mesa} to derive the full stellar parameters by combining our data. We also allowed the fitting of a linear slope on the original RVs as part of our joint analysis.

EXOFASTv2 considers the chains to be well-mixed when the Gelman-Rubin statistic \citep{gelman2004bayesian,gelman1992inference,ford2006improving} decreases below 1.01. The Gelman-Rubin statistic describes how similar the chains are; a value under 1.01 shows the chains are well-mixed. The list of our priors and final median values of the posterior distributions, and
their 1 $\sigma$ confidence intervals of full system parameters are reported in table \ref{tab:hd207897.}.

\begin{figure}[h!]
    \begin{tabular}{@{}c@{}}
	\includegraphics[width=0.9\columnwidth]{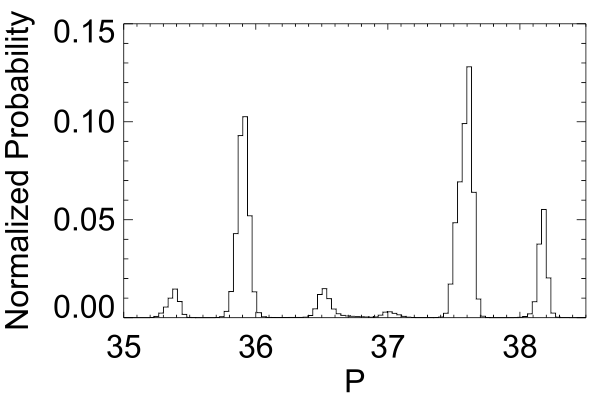}\\
	\includegraphics[width=0.9\columnwidth]{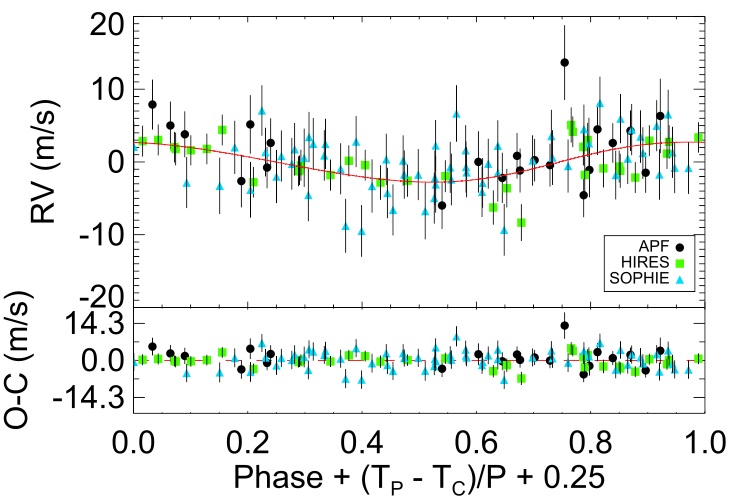}
	\end{tabular}
	\caption{\emph{Top:} Bimodality in the posterior distribution for the stellar activity periods. \emph{ Bottom:} Solution of the two-Keplerian model on the activity signal at 37.6 d with the initial period at 16.2 d.}
	\label{fig:37_keplerian} 
\end{figure}

As we showed in Sect \ref{Stellar rotation and activity}, HD207897 has a moderate stellar activity and the signals at 35.9 and 37.6 d are likely to be due to the stellar rotation periods. Since these signals affect slightly the mass estimate of the planet, we considered them as an additional keplerian fit in our global analysis. A more ideal solution to take activity into account is de-trending the RVs using activity indicators (e.g. S-index, FWHM), but currently, it is not possible with EXOFASTv2. Furthermore, to de-trend we did not have the same activity indicators for all RVs. As an independent analysis, we used only SOPHIE data and trained GP against CRX, the results were completely in agreement with EXOFASTv2 global modeling. When using two Keplerian models in EXOFASTv2, we did not fix the period and let EXOFASTv2 find the best activity period between the 35.9 and 37.6 d signals. After EXOFASTv2 converged, we saw a bimodality in the posterior distribution for the stellar activity periods (see Fig. \ref{fig:37_keplerian}, top), so we present the final median posterior distribution values of both the most probable solutions in Table \ref{tab:hd207897.}. We also report their calculated probabilities based on the area of the posterior distributions. The probability of the most likely values is 54\% and that of the less likely values is 46\%.

HD207897 is a main-sequence K0 dwarf star. We found that its most likely values for: mass is $0.800^{+0.036}_{-0.030}$ $M_{\sun}$, radius is $0.779^{+0.019}_{-0.018}$ R$_{\sun}$ and T$_{eff}$ is $5070^{+60}_{-57}$ K. These values are in good agreement with the result of our stellar analysis in Sect. \ref{subsec:spectralanalysis} and also, the stellar parameters from Gaia Data Release 2, such as T$_{eff}$= $5052^{+114}_{-78}$ K and radius of $0.79^{+0.80}_{-0.75}$ R$_{\sun}$ \citep{brown2018gaia}

We also show that the planet has a period of 16.20 d, a radius of $2.5\pm0.08$ R$_{\mathrm{E}}$ and a mass of either $14.4\pm 1.6$  M$_{\mathrm{E}}$ or $15.9\pm1.6$ M$_{\mathrm{E}}$ with nearly equal probability. These two solutions correspond to two possibilities for the stellar activity period. Hence, the density is either $ 5.1\pm0.7$ g cm$^{-3}$ or $5.5^{+0.8}_{-0.7}$ g cm$^{-3}$. Both values are in agreement with each other inside their error bars.

The keplerian solution and transit model for HD207897 b on the most likely values are shown in Fig. \ref{fig:lightcurve} (bottom). Also, the keplerian model on the most likely activity period on 37.6 d is illustrated in Fig. \ref{fig:37_keplerian}.

\section{Internal structure}

In order to characterise the internal structure of HD207897 b, we have performed a Markov chain Monte Carlo (MCMC) Bayesian analysis \citep{Dorn15} using the interior composition model introduced in \cite{brugger17}, \cite{mousis20} and \cite{acuna2021}, which comprises three layers: a Fe-rich core, a silicate rich mantle and a water layer.  With an input equilibrium temperature of 637 K, assuming an albedo of zero, the irradiance HD207897 b received is enough to present vapour and supercritical phases if water is found on its surface. Therefore we coupled an atmosphere-interior model that calculates the surface conditions and the contribution of the atmosphere to the total radius.
\begin{figure}[h!]
\centering
\includegraphics[width=0.4\textwidth]{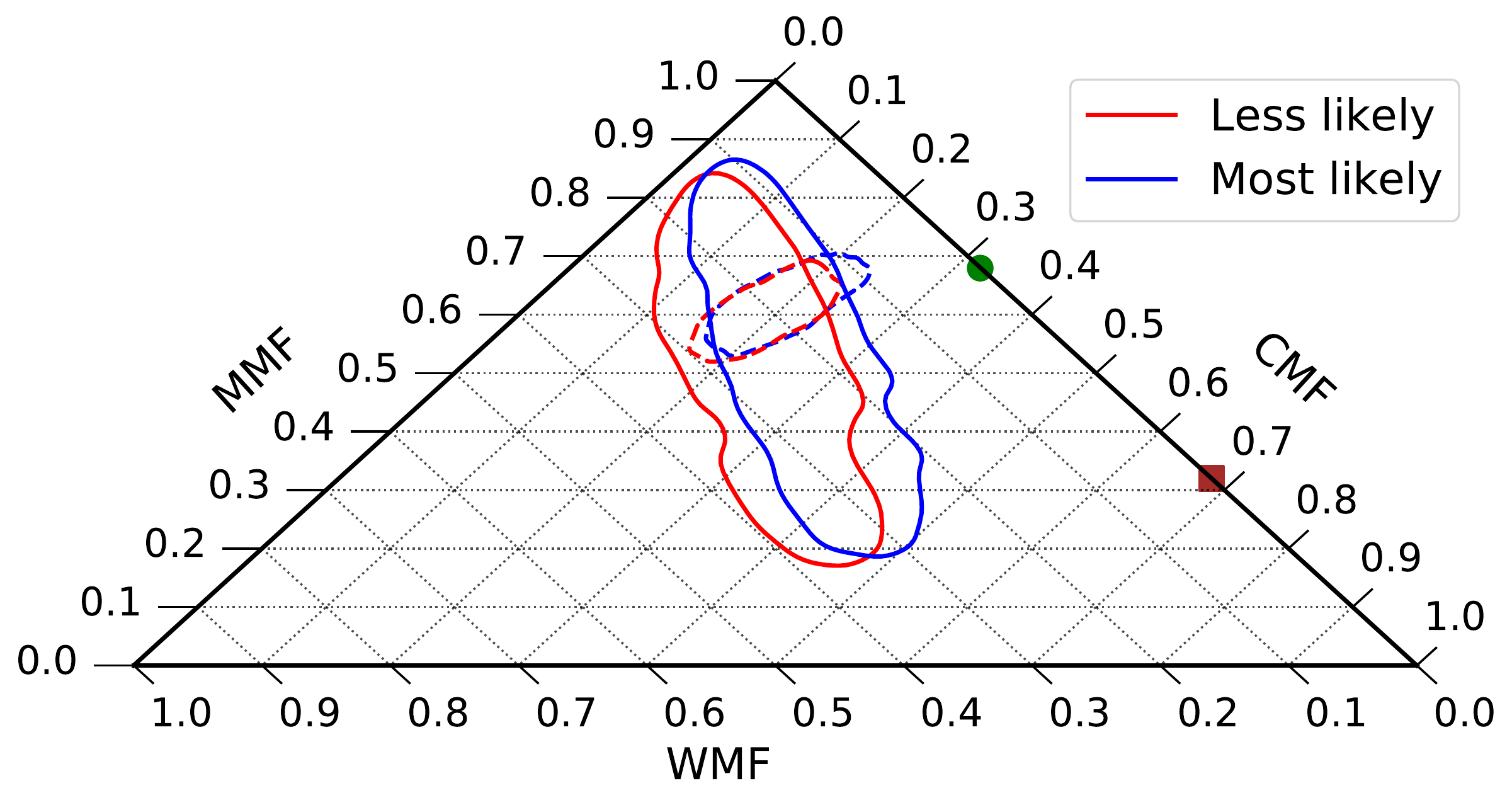}
\caption{2D, 1$\sigma$ confidence regions of HD207897 b for the mass estimation with activity period on 37.6 d (blue) and with activity period on 35.9 d (red). Solid lines indicate the confidence intervals of scenario 1, while dashed lines correspond to scenario 2, where the stellar abundances are also included as input data in the MCMC interior structure analysis. The mantle mass fraction (MMF) is defined as MMF = 1 - CMF - WMF. The green dot and brown square indicate the position of Earth and Mercury in the ternary diagram, respectively.}
\label{fig:ternary_comp}
\end{figure}

We considered two scenarios to obtain the interior structure of HD207897 b: scenario 1, where only the mass and radius of the planet are considered as inputs to the MCMC analysis (shown in Table \ref{tab:hd207897.}); and scenario 2, where the planetary mass and radius, and the stellar Fe/Si and Mg/Si mole ratios (see Table \ref{stellar parameters}) are the input data. To compute the Fe/Si and Mg/Si mole ratios with the stellar abundances, we follow the approach depicted in \cite{brugger17} and \cite{sotin07}, and obtain Fe/Si = 0.74$\pm$0.09 and Mg/Si = 1.11$\pm$0.20. The outputs of the MCMC analysis are the posterior distributions functions (PDF) of the core mass fraction (CMF), the water mass fraction (WMF) and the atmospheric parameters, which are the temperature at 300 bar, the planetary albedo and the atmospheric thickness from transit pressure to 300 bar. We assume a water-rich atmosphere. Table \ref{intrnal_struc} shows the 1D, 1$\sigma$ confidence intervals of the MCMC output parameters. 

In the most general case (scenario 1), up to 31\% of HD207897 b's mass can be in the form of a hydrosphere reaching supercritical or superionic phases at its base \citep{mazevet19} with the most likely value of the mass. This value increases slightly with the less likely value although both cases are consistent with a water-rich planet with a water mass fraction of 20 to 30 \%. The 1$\sigma$ confidence interval limits the maximum CMF to 0.50 (see Figure \ref{fig:ternary_comp}).
In addition, a Fe-depleted planet (CMF = 0) is possible, although unlikely. Assuming a pure silicate interior, the WMF is found to be 2.2 x $10^{-5}$ in HD207897 b, which corresponds to a pressure at the base of the hydrosphere of approximately 300 bar.

If we take into account the stellar abundances to constrain the planetary Fe/Si and Mg/Si mole ratios, the CMF is calculated as 0.21, a value lower than the Earth's CMF (0.32). In this scenario, the WMF would be below 0.16, implying that HD207897 b can be considered a water-rich planet. If the temperature and pressure of the hydrosphere are high enough to sustain a supercritical regime, then the atmosphere is extended and constitutes approximately 20\% of the planet's total radius.

\section{Discussion and Summary}

\begin{figure}
 	\includegraphics[width=0.48\textwidth]{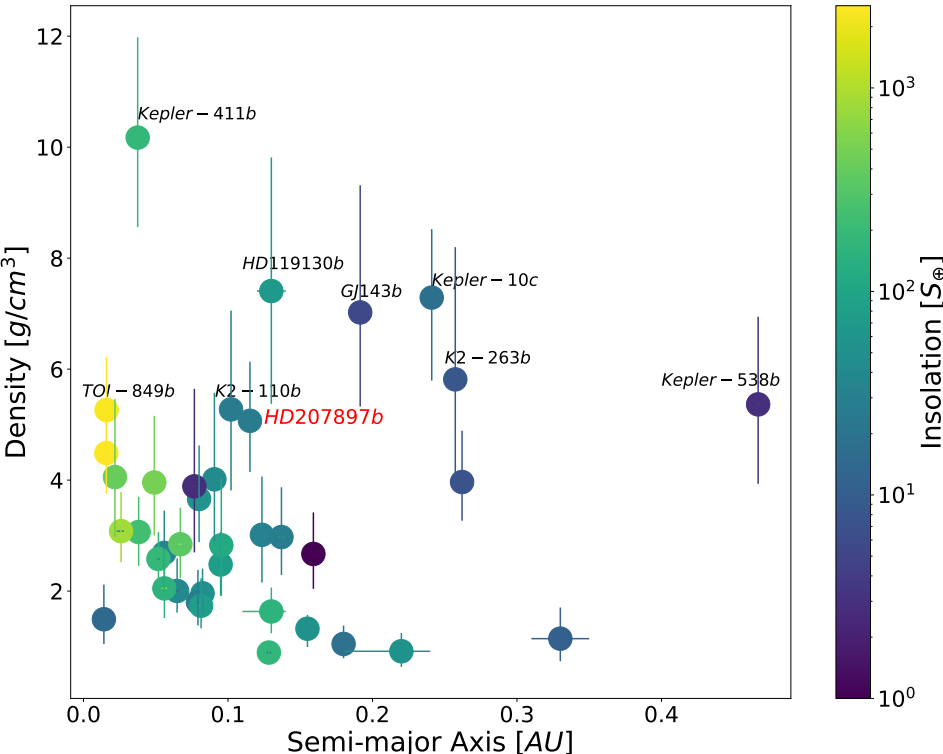}
	\caption{
           The density-semi major axis diagram for HD207897 b and other sub-Neptune sized planets (2 < R$_{\mathrm{E}}$ < 4) with known semi-major axis, luminosity and accurate mass and radius \citep{otegi2020revisited}. Dots are colored with planet insolation in Earth unit $ (S_{p}/S_{\bigoplus}= (L_{*}/L_{\sun})\times (AU/a_{p})^{2})$. The mentioned planet has a density of above 5 g cm$^{-3}$. }
	\label{fig:Mass-Raduis}
\end{figure}

 In this paper, we detected and characterized a sub-Neptune orbiting around HD207897, with a period of $16.202161\pm0.000083$ d, using TESS photometry data along with SOPHIE, HIRES, and APF RVs observations. We found the planet has a mass of $14.4\pm 1.6$ M$_{\mathrm{E}}$ with probability of 56 \% (or $15.9\pm1.6$ M$_{\mathrm{E}}$ with probability of 46 \% based on bimodal results of stellar activity 
  \begin{table*}
\centering
\begin{tabular}{lcccc}
\hline
\multicolumn{1}{c}{} & \multicolumn{2}{c}{Activity on 37.6 d} & \multicolumn{2}{c}{Activity on 35.9 d} \\
Parameter & Scenario 1 & Scenario 2 & Scenario 1 & Scenario 2 \\ \hline
Core mass fraction, CMF & 0.26$\pm$0.18 & 0.19$\pm$0.03 & 0.23$\pm$0.18 & 0.19$\pm$0.03 \\
Water mass fraction, WMF & 0.22$\pm$0.09 & 0.19$\pm$0.07 & 0.25$\pm$0.10 & 0.21$\pm$0.07 \\
Temperature at 300 bar, T$_{300}$ [K] & \multicolumn{2}{c}{2715$\pm$22} & 2742$\pm$24 & 2737$\pm$24 \\
Thickness at 300 bar, z$_{300}$ [km] & \multicolumn{2}{c}{234$\pm$22} & 265$\pm$26 & 257$\pm$25 \\
Albedo, a$_{p}$ & \multicolumn{4}{c}{0.26$\pm$0.01} \\
Core+Mantle radius, [R$_{p}$ units] & 0.72$\pm$0.08 & 0.75$\pm0.05$ & 0.70$\pm$0.08 & 0.74$\pm$0.05 \\ \hline
\end{tabular}
\caption{1$\sigma$ confidence intervals of the interior and atmosphere MCMC output parameters in the two different compositional scenarios (see text).}
\label{intrnal_struc}
\end{table*}
 
 \noindent
  period) and a radius of $2.5\pm0.08$ R$_{\mathrm{E}}$, which translates into a high density of $ 5.1\pm0.7$ g cm$^{-3}$ (or $5.5^{+0.8}_{-0.7}$ g cm$^{-3}$). We used the same mass and radii bounds as  \cite{otegi2020revisited} on \text NASA Exoplanet Data Archive\footnote{\url{https://exoplanetarchive.ipac.caltech.edu/}} (December 5th, 2020) and plotted in Fig \ref{fig:Mass-Raduis} all the sub-Neptune sized planets (2 < R/R$_{\mathrm{E}}$ < 4) with semi-major axis and luminosity determined. As shown in this plot, HD207897 b joins the group of dense sub-Neptunes such as HD 119130 b
 \citep[$\rho_P = 7.4^{+1.6}_{-1.5}$ g cm$^{-3}$;][]
{luque2018detection}, GJ143 b \citep[$\rho_P= 7^{+1.6}_{-1.3}$ g cm$^{-3}$;][] 
 {dragomir2019tess}, Kepler-10 c \citep[
 $\rho_P= 7.1 \pm 1$ g cm$^{-3}$;][] {dumusque2014kepler}, TOI-849 b \citep[$\rho_P= 5.2^{+0.7}_{-0.8}$ g cm$^{-3}$;][] {armstrong2020remnant}, Kepler-538 b \citep[$\rho_P= 5.4 \pm 1.3$ g cm$^{-3}$;][] {mayo201911}, Kepler-411 b \citep[$\rho_P= 9.9 \pm 1.3$ g cm$^{-3}$;][] {sun2019kepler}, K2-110 b \citep[$\rho_P= 5.2 \pm 1.2$ g cm$^{-3}$;][] {osborn2017k2}, and K2-263 b \citep[$\rho_P= 5.7^{+1.6}_{-1.4}$ g cm$^{-3}$;][] {mortier2018k2}. These planets are relatively close to their host stars and may have a similar formation history.

How can HD207897 b and other similar planets with such a high density exist at a close distance from their host star? One possibility could be that the planet has lost most of its volatile elements by evaporation, but for the case of HD207897 b with an orbital period of 16.20 d and receiving an incident flux of F=26.3  F$_{\mathrm{E}}$ it is not a satisfactory answer. Indeed, even if we consider an extreme evaporation process \citep{des2007diagram}, the mass-loss of the planet would be just 0.1 M$_{\mathrm{E}}$ during the entire lifetime of the star, which cannot account for its high density. HD207897 b is unlikely to have formed in situ. According to \cite{schlichting2014formation} the maximum isolation mass that can form at a distance of a = 0.12 AU is only $\sim$ 0.06 M$_{\mathrm{E}}$, assuming the minimum mass solar nebula (MMSN); a disk surface density $\sim$ 41 times larger than the solar nebula would be required to form a planet as massive as HD207897 b at this distance. Two possible scenarios can however be considered, both consistent with the MMSN: either the material from the outer region migrated and formed the planet HD207897 b \citep[e.g.][]{chatterjee2013inside} or the formation of HD207897 b occurred far out of the disc and the planet subsequently migrated to its current location \citep[e.g.][]{mcneil2010formation, kley2012planet}. The second scenario could have been triggered by another planet in this system. The hint of long-term trend on RVs allows for the presence of another planet. The high occurrence rate of long-period giant planets (mass > 0.3 M$_{\mathrm{J}}$) in the systems harboring small planets (planets with mass/radius between Earth and Neptune) \citep{zhu2018super, schlecker2020new} can also support such a scenario. In the case of HD 119130 b, \cite{luque2018detection} suggested a migration scenario triggered by other planets to explain the linear drift in their RV data. However, more photometry and radial velocity observations are still needed to understand the HD207897 planetary system. 
 
Further, the brightness (K= 6.3 mag) of HD207897, its relatively small stellar radius, and its quite nearby distance (28 pc) would make HD207897 b a good target for atmospheric characterization. Similarly, such a close distance and brightness for a transiting planet host, make HD207897 b an excellent target for studying the architecture of the system through ground-based observations. For example, measuring the host star spin-orbit alignment (obliquity) using the Rossiter-McLaughlin \citep[RM,][]{rossiter1924detection,mclaughlin1924some} anomaly can provide us with important information about planetary migration and evolution.  
%\newpage
\onecolumn
\providecommand{\bjdtdb}{\ensuremath{\rm {BJD_{TDB}}}}
\providecommand{\feh}{\ensuremath{\left[{\rm Fe}/{\rm H}\right]}}
\providecommand{\teff}{\ensuremath{T_{\rm eff}}}
\providecommand{\teq}{\ensuremath{T_{\rm eq}}}
\providecommand{\ecosw}{\ensuremath{e\cos{\omega_*}}}
\providecommand{\esinw}{\ensuremath{e\sin{\omega_*}}}
\providecommand{\msun}{\ensuremath{\,M_\Sun}}
\providecommand{\rsun}{\ensuremath{\,R_\Sun}}
\providecommand{\lsun}{\ensuremath{\,L_\Sun}}
\providecommand{\mj}{\ensuremath{\,M_{\rm J}}}
\providecommand{\rj}{\ensuremath{\,R_{\rm J}}}
\providecommand{\me}{\ensuremath{\,M_{\rm E}}}
\providecommand{\re}{\ensuremath{\,R_{\rm E}}}
\providecommand{\fave}{\langle F \rangle}
\providecommand{\fluxcgs}{10$^9$ erg s$^{-1}$ cm$^{-2}$}
\onecolumn
%\startlong
\begin{deluxetable}{lcccccc}
\tablewidth{0.9\columnwidth}
\tabletypesize{\scriptsize}
\tablecaption{Median values and 68\% confidence interval for HD207897 b and its host star. Gaussian priors are presented by $\mathcal{N} (a,b)$ where a and b are mean and width value, respectively. Likewise, uniform prior is donated by $\mathcal{U} (c,d)$ and c and d are presenting bounds on the parameter. The two solutions correspond to two possibilities for the stellar activity period. Highlighted parameters are presenting more than 0.5 $\sigma$ difference between the two solutions. See Table 3 in \cite{eastman2019exofastv2} for a detailed description of all parameters.}
\tablehead{\colhead{~~~Parameter} & \colhead{Units} & \colhead{Model priors}& \colhead{Activity on 35.9 d (Prob. = 46\%)}&\colhead{Activity on 37.6 d (Prob. = 54\%)}}
\startdata
\smallskip\\\multicolumn{2}{l}{Stellar Parameters:}& && \smallskip\\
~~~~$M_{*}$\dotfill & Mass (M$_{\sun}$)\dotfill&- &$0.801^{+0.036}_{-0.031}$
& $0.800^{+0.036}_{-0.030}$\\
~~~~$R_{*}$\dotfill &Radius (R$_{\sun}$)\dotfill &-&$0.779^{+0.019}_{-0.018}$&$0.779^{+0.019}_{-0.018}$\\
~~~~$R_{*,SED}$\dotfill &Radius$^{1}$ (R$_{\sun}$)\dotfill &-&$0.7691^{+0.011}_{-0.0098}$&$0.7692^{+0.011}_{-0.0100}$\\
~~~~$L_{*}$\dotfill &Luminosity (l$_{\sun}$)\dotfill &-&$0.360^{+0.019}_{-0.014}$&$0.360^{+0.019}_{-0.014}$\\
~~~~$F_{Bol}$\dotfill &Bolometric Flux (cgs)\dotfill&- &$0.00000001440^{+0.00000000076}_{-0.00000000056}$&$0.00000001439^{+0.00000000077}_{-0.00000000055}$\\
~~~~$\rho_{*}$\dotfill &Density (cgs)\dotfill &-&$2.39^{+0.19}_{-0.18}$&$2.39^{+0.19}_{-0.18}$\\
~~~~$\log{g}$\dotfill &Surface gravity (cgs)\dotfill&-&$4.558^{+0.027}_{-0.025}$&$4.559^{+0.026}_{-0.025}$\\
~~~~$T_{\rm eff}$\dotfill &Effective Temperature (K)\dotfill&$\mathcal{N}$[5012,80]&$5071^{+59}_{-57}$&$5070^{+60}_{-57}$\\
~~~~$T_{\rm eff,SED}$\dotfill &Effective Temperature$^{1}$ (K)\dotfill &-&$5096^{+78}_{-54}$&$5096^{+78}_{-54}$\\
~~~~$[{\rm Fe/H}]$\dotfill &Metallicity (dex)\dotfill &$\mathcal{N}[-0.21,0.08]$&$-0.040^{+0.037}_{-0.040}$&$-0.042^{+0.038}_{-0.041}$\\
~~~~$[{\rm Fe/H}]_{0}$\dotfill &Initial Metallicity$^{2}$ \dotfill &-&$-0.028^{+0.050}_{-0.052}$&$-0.030^{+0.050}_{-0.053}$\\
~~~~$Age$\dotfill &Age (Gyr)\dotfill&-&$7.1^{+4.5}_{-4.4}$&$7.1\pm4.5$\\
~~~~$EEP$\dotfill &Equal Evolutionary Phase$^{3}$ \dotfill &-&$343^{+17}_{-26}$&$343^{+17}_{-27}$\\
~~~~$A_V$\dotfill &V-band extinction (mag)\dotfill &$\mathcal{U}$[0,0.3534]&$0.074^{+0.081}_{-0.052}$&$0.074^{+0.081}_{-0.052}$\\
~~~~$\sigma_{SED}$\dotfill &SED photometry error scaling \dotfill &-&$2.10^{+0.77}_{-0.47}$&$2.10^{+0.77}_{-0.48}$\\
~~~~$\varpi$\dotfill &Parallax (mas)\dotfill &$\mathcal{N}[35.345,0.047]$&$35.345^{+0.048}_{-0.047}$&$35.345\pm0.047$\\
~~~~$d$\dotfill &Distance (pc)\dotfill&-&$28.292\pm0.038$&$28.293\pm0.038$\\
~~~~$\dot{\gamma}$\dotfill &RV slope$^{4}$ (m/s/day)\dotfill &-&$-0.00078\pm0.00030$&$-0.00077^{+0.00027}_{-0.00028}$\\
\smallskip\\\multicolumn{2}{l}{Planetary Parameters:}&&&\smallskip\\
~~~~$P$\dotfill &Period (days)\dotfill&- &$16.202157\pm0.000085$&$16.202159^{+0.000085}_{-0.000083}$\\
~~~~$R_P$\dotfill &Radius (\re)\dotfill&- &$2.505^{+0.081}_{-0.077}$&$2.501^{+0.082}_{-0.078}$\\
~~~~$M_P$\dotfill &Mass (\me)\dotfill &-&\textbf{15.9}$\pm1.6$&\textbf{14.4}$\pm1.6$\\
~~~~$T_C$\dotfill &Time of conjunction$^{5}$ (\bjdtdb)\dotfill &-&$2458926.10942^{+0.00047}_{-0.00049}$&$2458926.10942^{+0.00047}_{-0.00048}$\\
~~~~$T_T$\dotfill &Time of minimum projected separation$^{6}$ (\bjdtdb)\dotfill &-&$2458926.10944^{+0.00047}_{-0.00049}$&$2458926.10943^{+0.00047}_{-0.00048}$\\
~~~~$T_0$\dotfill &Optimal conjunction Time$^{7}$ (\bjdtdb)\dotfill &-&$2458926.10942^{+0.00047}_{-0.00049}$&$2458926.10942^{+0.00047}_{-0.00048}$\\
~~~~$a$\dotfill &Semi-major axis (AU)\dotfill &-&$0.1164^{+0.0017}_{-0.0015}$&$0.1163^{+0.0017}_{-0.0015}$\\
~~~~$i$\dotfill &Inclination (Degrees)\dotfill &-&$88.755^{+0.067}_{-0.066}$&$88.757\pm0.067$\\
~~~~$e$\dotfill &Eccentricity \dotfill &-&$0.047^{+0.053}_{-0.033}$&$0.047^{+0.057}_{-0.033}$\\
~~~~$\omega_*$\dotfill &Argument of Periastron (Degrees)\dotfill &-&$-125^{+68}_{-95}$&$-100\pm110$\\
~~~~$T_{eq}$\dotfill &Equilibrium temperature$^{8}$ (K)\dotfill &-&$632.3^{+8.2}_{-7.0}$&$632.2^{+8.1}_{-7.0}$\\
~~~~$\tau_{\rm circ}$\dotfill &Tidal circularization timescale (Gyr)\dotfill &-&$38700^{+8900}_{-7900}$&$35300^{+8300}_{-7300}$\\
~~~~$K$\dotfill &RV semi-amplitude (m/s)\dotfill &-&\textbf{4.65$^{+0.45}_{-0.46}$}&\textbf{4.24$^{+0.45}_{-0.46}$}\\
~~~~$R_P/R_*$\dotfill &Radius of planet in stellar radii \dotfill& -&$0.02946^{+0.00048}_{-0.00044}$&$0.02943^{+0.00049}_{-0.00046}$\\
~~~~$a/R_*$\dotfill &Semi-major axis in stellar radii \dotfill &-&$32.12^{+0.85}_{-0.82}$&$32.13^{+0.85}_{-0.82}$\\
~~~~$\delta$\dotfill &Transit depth (fraction)\dotfill &-&$0.000868^{+0.000029}_{-0.000026}$&$0.000866^{+0.000029}_{-0.000027}$\\
~~~~$Depth$\dotfill &Flux decrement at mid transit \dotfill& -&$0.000868^{+0.000029}_{-0.000026}$&$0.000866^{+0.000029}_{-0.000027}$\\
~~~~$\tau$\dotfill &Ingress/egress transit duration (days)\dotfill &-&$0.00684^{+0.0010}_{-0.00072}$&$0.00678^{+0.0011}_{-0.00077}$\\
~~~~$T_{14}$\dotfill &Total transit duration (days)\dotfill &-&$0.1214^{+0.0014}_{-0.0013}$&$0.1213^{+0.0014}_{-0.0013}$\\
~~~~$T_{FWHM}$\dotfill &FWHM transit duration (days)\dotfill &-&$0.1145^{+0.0012}_{-0.0011}$&$0.1145^{+0.0012}_{-0.0011}$\\
~~~~$b$\dotfill &Transit Impact parameter \dotfill &-&$0.712^{+0.041}_{-0.039}$&$0.709^{+0.043}_{-0.042}$\\
~~~~$b_S$\dotfill &Eclipse impact parameter \dotfill &-&$0.687^{+0.035}_{-0.047}$&$0.689^{+0.036}_{-0.048}$\\
~~~~$\tau_S$\dotfill &Ingress/egress eclipse duration (days)\dotfill &-&$0.00640^{+0.00066}_{-0.00069}$&$0.00644^{+0.00070}_{-0.00071}$\\
~~~~$T_{S,14}$\dotfill &Total eclipse duration (days)\dotfill &-&$0.1208^{+0.0016}_{-0.0019}$&$0.1209^{+0.0018}_{-0.0019}$\\
~~~~$T_{S,FWHM}$\dotfill &FWHM eclipse duration (days)\dotfill &-&$0.1143\pm0.0015$&$0.1143\pm0.0016$\\
~~~~$\delta_{S,2.5\mu m}$\dotfill &Blackbody eclipse depth at 2.5$\mu$m (ppm)\dotfill &-&$0.205^{+0.026}_{-0.021}$&$0.204^{+0.026}_{-0.021}$\\
~~~~$\delta_{S,5.0\mu m}$\dotfill &Blackbody eclipse depth at 5.0$\mu$m (ppm)\dotfill &-&$7.09^{+0.50}_{-0.45}$&$7.07^{+0.50}_{-0.45}$\\
~~~~$\delta_{S,7.5\mu m}$\dotfill &Blackbody eclipse depth at 7.5$\mu$m (ppm)\dotfill &-&$20.2^{+1.1}_{-1.0}$&$20.2^{+1.1}_{-1.0}$\\
~~~~$\rho_P$\dotfill &Density (cgs)\dotfill &-&\textbf{5.52}$^{+0.82}_{-0.73}$&\textbf{5.05}$^{+0.77}_{-0.69}$\\
~~~~$logg_P$\dotfill &Surface gravity \dotfill &-&\textbf{3.393}$^{+0.051}_{-0.053}$&\textbf{3.353}$^{+0.053}_{-0.056}$\\
~~~~$\Theta$\dotfill &Safronov Number \dotfill &-&\textbf{0.0647}$^{+0.0067}_{-0.0066}$&\textbf{0.0589}$^{+0.0066}_{-0.0065}$\\
~~~~$\fave$\dotfill &Incident Flux (\fluxcgs)\dotfill &-&$0.0361^{+0.0019}_{-0.0016}$&$0.0361^{+0.0019}_{-0.0016}$\\
~~~~$T_P$\dotfill &Time of Periastron (\bjdtdb)\dotfill &-&$2458916.2^{+3.4}_{-4.3}$&$2458917.0^{+4.5}_{-5.4}$\\
~~~~$T_S$\dotfill &Time of eclipse (\bjdtdb)\dotfill &-&$2458934.08^{+0.25}_{-0.40}$&$2458934.18^{+0.30}_{-0.38}$\\
~~~~$T_A$\dotfill &Time of Ascending Node (\bjdtdb)\dotfill &-&$2458921.93^{+0.22}_{-0.42}$&$2458922.00^{+0.25}_{-0.42}$\\
~~~~$T_D$\dotfill &Time of Descending Node (\bjdtdb)\dotfill &-&$2458930.16^{+0.31}_{-0.26}$&$2458930.18^{+0.36}_{-0.27}$\\
~~~~$V_c/V_e$\dotfill & \dotfill &-&$1.010^{+0.067}_{-0.034}$&$1.007^{+0.069}_{-0.039}$\\
~~~~$e\cos{\omega_*}$\dotfill & \dotfill &-&$-0.012^{+0.024}_{-0.039}$&$-0.003^{+0.029}_{-0.037}$\\
~~~~$e\sin{\omega_*}$\dotfill & \dotfill &-&$-0.011^{+0.034}_{-0.065}$&$-0.007^{+0.039}_{-0.066}$\\
~~~~$M_P\sin i$\dotfill &Minimum mass (\me)\dotfill &-&\textbf{15.8}$\pm1.6$&\textbf{14.4}$\pm1.6$\\
~~~~$M_P/M_*$\dotfill &Mass ratio \dotfill &-&\textbf{0.0000594}$\pm0.0000058$&\textbf{0.0000540}$\pm0.0000058$\\
~~~~$d/R_*$\dotfill &Separation at mid transit \dotfill &-&$32.6^{+2.1}_{-1.6}$&$32.5^{+2.2}_{-1.7}$\\
~~~~$P_T$\dotfill &A priori non-grazing transit prob \dotfill &-&$0.0298^{+0.0015}_{-0.0018}$&$0.0299^{+0.0017}_{-0.0019}$\\
~~~~$P_{T,G}$\dotfill &A priori transit prob \dotfill &-&$0.0316^{+0.0016}_{-0.0019}$&$0.0317^{+0.0018}_{-0.0020}$\\
~~~~$P_S$\dotfill &A priori non-grazing eclipse prob \dotfill &-&$0.0307^{+0.0021}_{-0.0015}$&$0.0306^{+0.0022}_{-0.0016}$\\
~~~~$P_{S,G}$\dotfill &A priori eclipse prob \dotfill &-&$0.0326^{+0.0023}_{-0.0016}$&$0.0325^{+0.0023}_{-0.0017}$\\
\smallskip\\\multicolumn{2}{l}{Stellar activity parameters:}&\smallskip\\
~~~~$P_{Activity}$\dotfill &Period (days)\dotfill &-&\textbf{35.923}$^{+0.069}_{-0.067}$&\textbf{37.627}$^{+0.55}_{-0.076}$\\
~~~~$T_{C, activity}$\dotfill &Time of conjunction$^{5}$ (\bjdtdb)\dotfill &-&\textbf{2458917.1}$\pm2.2$&\textbf{2458910.7}$^{+2.3}_{-2.6}$\\
\\
\\
\\
\\
~~~~$e_{Activity}$\dotfill &Eccentricity \dotfill &-&$0.14^{+0.13}_{-0.10}$&$0.15^{+0.13}_{-0.11}$\\
~~~~$\omega_{*, activity}$\dotfill &Argument of Periastron (Degrees)\dotfill &-&\textbf{-127}$^{+98}_{-96}$&\textbf{33}$^{+94}_{-92}$\\
~~~~$K_{Activity}$\dotfill &RV semi-amplitude (m/s)\dotfill &-&$2.38^{+0.51}_{-0.52}$&$2.25^{+0.50}_{-0.49}$\\
\smallskip\\\multicolumn{2}{l}{Telescope Parameters for SOPHIE:}&\smallskip\\
~~~~$\gamma_{\rm rel}$\dotfill &Relative RV Offset$^{4}$ (m/s)\dotfill &-&$-6327.36\pm0.56$&$-6327.53\pm0.52$\\
~~~~$\sigma_J$\dotfill &RV Jitter (m/s)\dotfill &-&$3.16^{+0.42}_{-0.37}$&$3.05^{+0.41}_{-0.37}$\\
~~~~$\sigma_J^2$\dotfill &RV Jitter Variance \dotfill &-&$10.0^{+2.8}_{-2.2}$&$9.3^{+2.7}_{-2.1}$\\
\smallskip\\\multicolumn{2}{l}{Telescope
Parameters for APF:}&&&\smallskip\\
~~~~$\gamma_{\rm rel}$\dotfill &Relative RV Offset$^{4}$ (m/s)\dotfill &-&$3.1\pm1.1$&$2.9^{+1.1}_{-1.0}$\\
~~~~$\sigma_J$\dotfill &RV Jitter (m/s)\dotfill &-&$1.6^{+1.2}_{-1.6}$&$1.4^{+1.3}_{-1.4}$\\
~~~~$\sigma_J^2$\dotfill &RV Jitter Variance \dotfill &-&$2.7^{+5.5}_{-3.4}$&$2.0^{+5.4}_{-3.2}$\\
\smallskip\\\multicolumn{2}{l}{Telescope
Parameters for HIRES:}&&\smallskip\\
~~~~$\gamma_{\rm rel}$\dotfill &Relative RV Offset$^{4}$ (m/s)\dotfill &-&$1.61^{+0.83}_{-0.86}$&$1.81^{+0.81}_{-0.76}$\\
~~~~$\sigma_J$\dotfill &RV Jitter (m/s)\dotfill &-&\textbf{2.15}$^{+0.46}_{-0.38}$&\textbf{2.42}$^{+0.49}_{-0.40}$\\
~~~~$\sigma_J^2$\dotfill &RV Jitter Variance \dotfill &-&$4.6^{+2.2}_{-1.5}$&$5.9^{+2.6}_{-1.8}$\\
\\
\smallskip\\\multicolumn{2}{l}{Wavelength Parameters:}&&TESS&TESS\smallskip\\
~~~~$u_{1}$\dotfill &linear limb-darkening coeff \dotfill &-&$0.386\pm0.046$&$0.385^{+0.046}_{-0.047}$\\
~~~~$u_{2}$\dotfill &quadratic limb-darkening coeff \dotfill &-&$0.188\pm0.048$&$0.189^{+0.047}_{-0.048}$\\
\smallskip\\\multicolumn{2}{l}{Transit Parameters:}&&TESS UT 2019-10-10 (TESS)&TESS UT 2019-10-10 (TESS)\smallskip\\
~~~~$\sigma^{2}$\dotfill &Added Variance \dotfill &-&$0.0000000244\pm0.0000000011$&$0.0000000244\pm0.0000000011$\\
~~~~$F_0$\dotfill &Baseline flux \dotfill &-&$0.9999796^{+0.0000022}_{-0.0000023}$&$0.9999796^{+0.0000022}_{-0.0000023}$\\
\enddata
\label{tab:hd207897.}
\tablenotetext{1}{This value ignores the systematic error and is for reference only}
\tablenotetext{2}{The metallicity of the star at birth}
\tablenotetext{3}{Corresponds to static points in a star's evolutionary history. See \S2 in \cite{dotter2016mesa}}
\tablenotetext{4}{Reference epoch = 2456438.359500}
\tablenotetext{5}{Time of conjunction is commonly reported as the "transit time"}
\tablenotetext{6}{Time of minimum projected separation is a more correct "transit time"}
\tablenotetext{7}{Optimal time of conjunction minimizes the covariance between $T_C$ and Period}
\tablenotetext{8}{Assumes no albedo and perfect redistribution}
\end{deluxetable}

\twocolumn

%%%%%%%%%%%%%%%%%%%%%%%%%%%%%%%%%%%%%%%%%

\begin{acknowledgements}
      We warmly thank the OHP staff for their support on the observations. X.B., I.B. and T.F. received funding from the French Programme National de Physique Stellaire (PNPS) and the Programme National de Planétologie (PNP) of CNRS (INSU). We also acknowledge the financial support of French embassy in Tehran. 
      \\
      N. H acknowledges F. Vakili for his constant academic and administrative support. N.H warmly thank Ph. Stee, J. L. Beuzit, E. T. Givenchy for all their help. N.H also thanks Jason D. Eastman and Michael Hippke who have written EXOFASTv2 and TLS package, respectively, for their guides. This publication makes use of The Data \& Analysis Center for Exoplanets (DACE), which is a facility based at the University of Geneva (CH) dedicated to extrasolar planets data visualisation, exchange and analysis. DACE is a platform of the Swiss National Centre of Competence in Research (NCCR) PlanetS, federating the Swiss expertise in Exoplanet research. The DACE platform is available at \url{https://dace.unige.ch}.
      \\
      J.L-B. acknowledges financial support received from "la Caixa" Foundation (ID 100010434) and from the European Union's Horizon 2020 research and innovation programme under the Marie Skłodowska-Curie grant agreement No 847648, with fellowship code LCF/BQ/PI20/11760023. This work has been carried out in the frame of the National Centre for Competence in Research "Planets" supported by the Swiss National Science Foundation (SNSF). This project has received funding from the European Research Council (ERC) under the European Union's Horizon 2020 research and innovation programme (project {\sc Spice Dune}, grant agreement No 947634). This work was supported by FCT - Funda\c{c}\~ao para a Ci\^encia e Tecnologia (FCT) through national funds and by FEDER through COMPETE2020 - Programa Operacional Competitividade e Internacionaliza\c{c}\~ao by these grants: UID/FIS/04434/2019; UIDB/04434/2020; UIDP/04434/2020; PTDC/FIS-AST/32113/2017 \& POCI-01-0145-FEDER-032113; PTDC/FIS-AST/28953/2017 \& POCI-01-0145-FEDER-028953. V.A., E.D.M, N.C.S., and S.G.S. also acknowledge the support from FCT through Investigador FCT contracts nr.  IF/00650/2015/CP1273/CT0001, IF/00849/2015/CP1273/CT0003, IF/00169/2012/CP0150/CT0002, and IF/00028/2014/CP1215/CT0002, respectively, and POPH/FSE (EC) by FEDER funding through the program "Programa Operacional de Factores de Competitividade - COMPETE".
      
      \\
      O.D.S.D. is supported in the form of work contract (DL 57/2016/CP1364/CT0004) funded by FCT. AC et PC acknowledge funding from the French National Research Agency (ANR) under contract number ANR-18-CE31-0019 (SPlaSH). N. A.D. acknowledges the support of FONDECYT project 3180063.SH acknowledges CNES funding through the grant 837319. XD and GG acknowledge funding in the framework of the Investissements d'Avenir program (ANR-15-IDEX-02), through the funding of the ”Origin of Life” project of the Univ. Grenoble-Alpes. 
      \\
      This work was supported by Fundação para a Ciência e a Tecnologia (FCT) and Fundo Europeu de Desenvolvimento Regional (FEDER) via COMPETE2020 through the research grants UIDB/04434/2020, UIDP/04434/2020, PTDC/FIS-AST/32113/2017 \& POCI-01-0145-FEDER-032113, PTDC/FIS-AST/28953/2017 \& POCI-01-0145-FEDER-028953. O.D.S.D. is supported in the form of work contract (DL 57/2016/CP1364/CT0004) funded by FCT.
      
      \\
      Resources supporting this work were provided by the NASA High-End Computing (HEC) Program through the NASA Advanced Supercomputing (NAS) Division at Ames Research Center for the production of the SPOC data products. Funding for the TESS mission is provided by NASA's Science Mission directorate. We acknowledge the use of public TESS Alert data from pipelines at the TESS Science Office and at the TESS Science Processing Operations Center. This research has made use of the Exoplanet Follow-up Observation Program website, which is operated by the California Institute of Technology, under contract with the National Aeronautics and Space Administration under the Exoplanet Exploration Program. This paper includes data collected by the TESS mission that are publicly available from the Mikulski Archive for Space Telescopes (MAST).
      \\
      D. D acknowledges support from the TESS Guest Investigator Program grant $80NSSC19K1727$ and NASA Exoplanet Research Program grant $18-2XRP18_2-0136$. T.D acknowledges support from MIT's Kavli Institute as a Kavli postdoctoral fellow. E.A.P. acknowledges the support of the Alfred P. Sloan Foundation. L.M.W. is supported by the Beatrice Watson Parrent Fellowship and NASA ADAP Grant 80NSSC19K0597. A.C. is supported by the NSF Graduate Research Fellowship, grant No. DGE 1842402. D.H. acknowledges support from the Alfred P. Sloan Foundation, the National Aeronautics and Space Administration (80NSSC19K0379), and the National Science Foundation (AST-1717000). I.J.M.C. acknowledges support from the NSF through grant AST-1824644. P.D. acknowledges support from a National Science Foundation Astronomy and Astrophysics Postdoctoral Fellowship under award AST-1903811. A.B. is supported by the NSF Graduate Research Fellowship, grant No. DGE 1745301. R.A.R. is supported by the NSF Graduate Research Fellowship, grant No. DGE 1745301. C. D. D. acknowledges the support of the Hellman Family Faculty Fund, the Alfred P. Sloan Foundation, the David \& Lucile Packard Foundation, and the National Aeronautics and Space Administration via the TESS Guest Investigator Program (80NSSC18K1583).  J.M.A.M. is supported by the NSF Graduate Research Fellowship, grant No. DGE-1842400. J.M.A.M. and P.C. acknowledges the LSSTC Data Science Fellowship Program, which is funded by LSSTC, NSF Cybertraining Grant No. 1829740, the Brinson Foundation, and the Moore Foundation; their participation in the program has benefited this work. 
      % http://www.not.iac.es/news/publications/
     This paper is partially based on observations made with the Nordic Optical Telescope, operated by the Nordic Optical Telescope Scientific Association at the Observatorio del Roque de los Muchachos, La Palma, Spain, of the Instituto de Astrofisica de Canarias.

\end{acknowledgements}

%-------------------------------------------------------------------
%\bibliography{biblio}
%\bibliographystyle{unsrt}
\bibliography{mtw}

\begin{thebibliography}{122}
\expandafter\ifx\csname natexlab\endcsname\relax\def\natexlab#1{#1}\fi

\bibitem[{{Acu{\~n}a} {et~al.}(2021){Acu{\~n}a}, {Deleuil}, {Mousis}, {Marcq},
  {Levesque}, \& {Aguichine}}]{acuna2021}
{Acu{\~n}a}, L., {Deleuil}, M., {Mousis}, O., {et~al.} 2021, arXiv e-prints,
  arXiv:2101.08172

\bibitem[{{Adibekyan} {et~al.}(2015){Adibekyan}, {Figueira}, {Santos}, {Sousa},
  {Faria}, {Delgado-Mena}, {Oshagh}, {Tsantaki}, {Hakobyan}, {Gonz{\'a}lez
  Hern{\'a}ndez}, {Su{\'a}rez-Andr{\'e}s}, \& {Israelian}}]{Adibekyan-15}
{Adibekyan}, V., {Figueira}, P., {Santos}, N.~C., {et~al.} 2015, \aap, 583, A94

\bibitem[{{Adibekyan} {et~al.}(2012){Adibekyan}, {Sousa}, {Santos}, {Delgado
  Mena}, {Gonz{\'a}lez Hern{\'a}ndez}, {Israelian}, {Mayor}, \&
  {Khachatryan}}]{Adibekyan-12}
{Adibekyan}, V.~Z., {Sousa}, S.~G., {Santos}, N.~C., {et~al.} 2012, \aap, 545,
  A32

\bibitem[{Aller {et~al.}(2020)Aller, Lillo-Box, Jones, Miranda, \&
  Forteza}]{aller2020planetary}
Aller, A., Lillo-Box, J., Jones, D., Miranda, L.~F., \& Forteza, S.~B. 2020,
  Astronomy \& Astrophysics, 635, A128

\bibitem[{Angus {et~al.}(2016)Angus, Foreman-Mackey, \&
  Johnson}]{angus2016systematics}
Angus, R., Foreman-Mackey, D., \& Johnson, J.~A. 2016, The Astrophysical
  Journal, 818, 109

\bibitem[{Armstrong {et~al.}(2020)Armstrong, Lopez, Adibekyan, Booth, Bryant,
  Collins, Deleuil, Emsenhuber, Huang, King, {et~al.}}]{armstrong2020remnant}
Armstrong, D.~J., Lopez, T.~A., Adibekyan, V., {et~al.} 2020, Nature, 583, 39

\bibitem[{Baliunas {et~al.}(1995)Baliunas, Donahue, Soon, Horne, Frazer,
  Woodard-Eklund, Bradford, Rao, Wilson, Zhang,
  {et~al.}}]{baliunas1995chromospheric}
Baliunas, S.~{\'a}., Donahue, R., Soon, W., {et~al.} 1995, The Astrophysical
  Journal, 438, 269

\bibitem[{Baluev(2008)}]{baluev2008assessing}
Baluev, R.~V. 2008, Monthly Notices of the Royal Astronomical Society, 385,
  1279

\bibitem[{{Bertran de Lis} {et~al.}(2015){Bertran de Lis}, {Delgado Mena},
  {Adibekyan}, {Santos}, \& {Sousa}}]{Bertrandelis-15}
{Bertran de Lis}, S., {Delgado Mena}, E., {Adibekyan}, V.~Z., {Santos}, N.~C.,
  \& {Sousa}, S.~G. 2015, \aap, 576, A89

\bibitem[{Bhatti {et~al.}(2020)Bhatti, Bouma, Joshua, John, \&
  Price-Whelan}]{astrobase}
Bhatti, W., Bouma, L., Joshua, John, \& Price-Whelan, A. 2020,
  waqasbhatti/astrobase: astrobase v0.5.0

\bibitem[{Boisse {et~al.}(2011)Boisse, Bouchy, H{\'e}brard, Bonfils, Santos, \&
  Vauclair}]{boisse2011disentangling}
Boisse, I., Bouchy, F., H{\'e}brard, G., {et~al.} 2011, Astronomy \&
  Astrophysics, 528, A4

\bibitem[{Borucki {et~al.}(2010)Borucki, Koch, Basri, Batalha, Brown, Caldwell,
  Caldwell, Christensen-Dalsgaard, Cochran, DeVore,
  {et~al.}}]{borucki2010kepler}
Borucki, W.~J., Koch, D., Basri, G., {et~al.} 2010, Science, 327, 977

\bibitem[{Bouchy {et~al.}(2013)Bouchy, D{\'\i}az, H{\'e}brard, Arnold, Boisse,
  Delfosse, Perruchot, \& Santerne}]{bouchy2013sophie+}
Bouchy, F., D{\'\i}az, R., H{\'e}brard, G., {et~al.} 2013, Astronomy \&
  Astrophysics, 549, A49

\bibitem[{Bouchy {et~al.}(2009{\natexlab{a}})Bouchy, H{\'e}brard, Udry,
  Delfosse, Boisse, Desort, Bonfils, Eggenberger, Ehrenreich, Forveille,
  {et~al.}}]{bouchy2009sophie}
Bouchy, F., H{\'e}brard, G., Udry, S., {et~al.} 2009{\natexlab{a}}, Astronomy
  \& Astrophysics, 505, 853

\bibitem[{Bouchy {et~al.}(2009{\natexlab{b}})Bouchy, H{\'e}brard, Udry,
  Delfosse, Boisse, Desort, Bonfils, Eggenberger, Ehrenreich, Forveille,
  {et~al.}}]{bouchy2009}
Bouchy, F., H{\'e}brard, G., Udry, S., {et~al.} 2009{\natexlab{b}}, Astronomy
  \& Astrophysics, 505, 853

\bibitem[{Bouchy {et~al.}(2008)Bouchy, Moutou, Queloz,
  {et~al.}}]{bouchy2008radial}
Bouchy, F., Moutou, C., Queloz, D., {et~al.} 2008, Proceedings of the
  International Astronomical Union, 4, 129

\bibitem[{Brewer {et~al.}(2016)Brewer, Fischer, Valenti, \&
  Piskunov}]{brewer2016spectral}
Brewer, J.~M., Fischer, D.~A., Valenti, J.~A., \& Piskunov, N. 2016, The
  Astrophysical Journal Supplement Series, 225, 32

\bibitem[{Brown {et~al.}(2018)Brown, Vallenari, Prusti, De~Bruijne, Babusiaux,
  Bailer-Jones, Biermann, Evans, Eyer, Jansen, {et~al.}}]{brown2018gaia}
Brown, A., Vallenari, A., Prusti, T., {et~al.} 2018, Astronomy \& astrophysics,
  616, A1

\bibitem[{{Brugger} {et~al.}(2017){Brugger}, {Mousis}, {Deleuil}, \&
  {Deschamps}}]{brugger17}
{Brugger}, B., {Mousis}, O., {Deleuil}, M., \& {Deschamps}, F. 2017, \apj, 850,
  93

\bibitem[{{Buchhave} {et~al.}(2010){Buchhave}, {Bakos}, {Hartman}, {Torres},
  {Kov{\'a}cs}, {Latham}, {Noyes}, {Esquerdo}, {Everett}, {Howard}, {Marcy},
  {Fischer}, {Johnson}, {Andersen}, {F{\H{u}}r{\'e}sz}, {Perumpilly},
  {Sasselov}, {Stefanik}, {B{\'e}ky}, {L{\'a}z{\'a}r}, {Papp}, \&
  {S{\'a}ri}}]{Buchhave2010}
{Buchhave}, L.~A., {Bakos}, G.~{\'A}., {Hartman}, J.~D., {et~al.} 2010, \apj,
  720, 1118

\bibitem[{{Buchhave} {et~al.}(2014){Buchhave}, {Bizzarro}, {Latham},
  {Sasselov}, {Cochran}, {Endl}, {Isaacson}, {Juncher}, \&
  {Marcy}}]{Buchhave2014}
{Buchhave}, L.~A., {Bizzarro}, M., {Latham}, D.~W., {et~al.} 2014, \nat, 509,
  593

\bibitem[{{Buchhave} {et~al.}(2012){Buchhave}, {Latham}, {Johansen},
  {Bizzarro}, {Torres}, {Rowe}, {Batalha}, {Borucki}, {Brugamyer}, {Caldwell},
  {Bryson}, {Ciardi}, {Cochran}, {Endl}, {Esquerdo}, {Ford}, {Geary},
  {Gilliland}, {Hansen}, {Isaacson}, {Laird}, {Lucas}, {Marcy}, {Morse},
  {Robertson}, {Shporer}, {Stefanik}, {Still}, \& {Quinn}}]{Buchhave2012}
{Buchhave}, L.~A., {Latham}, D.~W., {Johansen}, A., {et~al.} 2012, \nat, 486,
  375

\bibitem[{Burt {et~al.}(2014)Burt, Hanson, Rivera, Holden, Vogt, Butler,
  Arriagada, \& Laughlin}]{burt2014achieving}
Burt, J., Hanson, R., Rivera, E., {et~al.} 2014, in Software and
  Cyberinfrastructure for Astronomy III, Vol. 9152, International Society for
  Optics and Photonics, 915211

\bibitem[{Butler {et~al.}(1996)Butler, Marcy, Williams, McCarthy, Dosanjh, \&
  Vogt}]{butler1996attaining}
Butler, R.~P., Marcy, G.~W., Williams, E., {et~al.} 1996, Publications of the
  Astronomical Society of the Pacific, 108, 500

\bibitem[{Chatterjee \& Tan(2013)}]{chatterjee2013inside}
Chatterjee, S. \& Tan, J.~C. 2013, The Astrophysical Journal, 780, 53

\bibitem[{Choi {et~al.}(2016)Choi, Dotter, Conroy, Cantiello, Paxton, \&
  Johnson}]{choi2016mesa}
Choi, J., Dotter, A., Conroy, C., {et~al.} 2016, The Astrophysical Journal,
  823, 102

\bibitem[{Claret(2017)}]{claret2017limb}
Claret, A. 2017, Astronomy \& Astrophysics, 600, A30

\bibitem[{Courcol {et~al.}(2015)Courcol, Bouchy, Pepe, Santerne, Delfosse,
  Arnold, Astudillo-Defru, Boisse, Bonfils, Borgniet,
  {et~al.}}]{courcol2015sophie}
Courcol, B., Bouchy, F., Pepe, F., {et~al.} 2015, Astronomy \& Astrophysics,
  581, A38

\bibitem[{{Delgado Mena} {et~al.}(2010){Delgado Mena}, {Israelian},
  {Gonz{\'a}lez Hern{\'a}ndez}, {Bond}, {Santos}, {Udry}, \&
  {Mayor}}]{Delgado-10}
{Delgado Mena}, E., {Israelian}, G., {Gonz{\'a}lez Hern{\'a}ndez}, J.~I.,
  {et~al.} 2010, \apj, 725, 2349

\bibitem[{{Delgado Mena} {et~al.}(2017){Delgado Mena}, {Tsantaki}, {Adibekyan},
  {Sousa}, {Santos}, {Gonz{\'a}lez Hern{\'a}ndez}, \& {Israelian}}]{Delgado-17}
{Delgado Mena}, E., {Tsantaki}, M., {Adibekyan}, V.~Z., {et~al.} 2017, \aap,
  606, A94

\bibitem[{Delisle {et~al.}(2016)Delisle, S{\'e}gransan, Buchschacher, \&
  Alesina}]{delisle2016analytical}
Delisle, J.-B., S{\'e}gransan, D., Buchschacher, N., \& Alesina, F. 2016,
  Astronomy \& Astrophysics, 590, A134

\bibitem[{{Dorn} {et~al.}(2015){Dorn}, {Khan}, {Heng}, {Connolly}, {Alibert},
  {Benz}, \& {Tackley}}]{Dorn15}
{Dorn}, C., {Khan}, A., {Heng}, K., {et~al.} 2015, \aap, 577, A83

\bibitem[{Dotter(2016)}]{dotter2016mesa}
Dotter, A. 2016, The Astrophysical Journal Supplement Series, 222, 8

\bibitem[{Dragomir {et~al.}(2019)Dragomir, Teske, G{\"u}nther, S{\'e}gransan,
  Burt, Huang, Vanderburg, Matthews, Dumusque, Stassun,
  {et~al.}}]{dragomir2019tess}
Dragomir, D., Teske, J., G{\"u}nther, M.~N., {et~al.} 2019, The Astrophysical
  Journal Letters, 875, L7

\bibitem[{Dumusque {et~al.}(2014)Dumusque, Bonomo, Haywood, Malavolta,
  Segransan, Buchhave, Cameron, Latham, Molinari, Pepe,
  {et~al.}}]{dumusque2014kepler}
Dumusque, X., Bonomo, A.~S., Haywood, R.~D., {et~al.} 2014, The Astrophysical
  Journal, 789, 154

\bibitem[{Eastman(2017)}]{eastman2017exofastv2}
Eastman, J. 2017, Astrophysics Source Code Library

\bibitem[{Eastman {et~al.}(2013)Eastman, Gaudi, \& Agol}]{eastman2013exofast}
Eastman, J., Gaudi, B.~S., \& Agol, E. 2013, Publications of the Astronomical
  Society of the Pacific, 125, 83

\bibitem[{Eastman {et~al.}(2019)Eastman, Rodriguez, Agol, Stassun, Beatty,
  Vanderburg, Gaudi, Collins, \& Luger}]{eastman2019exofastv2}
Eastman, J.~D., Rodriguez, J.~E., Agol, E., {et~al.} 2019, arXiv preprint
  arXiv:1907.09480

\bibitem[{F\H{u}r\'esz(2008)}]{gaborthesis}
F\H{u}r\'esz, G. 2008, PhD thesis, University of Szeged, Hungary

\bibitem[{Findeisen {et~al.}(2011)Findeisen, Hillenbrand, \&
  Soderblom}]{findeisen2011stellar}
Findeisen, K., Hillenbrand, L., \& Soderblom, D. 2011, The Astronomical
  Journal, 142, 23

\bibitem[{Ford(2006)}]{ford2006improving}
Ford, E.~B. 2006, The Astrophysical Journal, 642, 505

\bibitem[{Fulton {et~al.}(2017)Fulton, Petigura, Howard, Isaacson, Marcy,
  Cargile, Hebb, Weiss, Johnson, Morton, {et~al.}}]{fulton2017california}
Fulton, B.~J., Petigura, E.~A., Howard, A.~W., {et~al.} 2017, The Astronomical
  Journal, 154, 109

\bibitem[{Fulton {et~al.}(2015)Fulton, Weiss, Sinukoff, Isaacson, Howard,
  Marcy, Henry, Holden, \& Kibrick}]{fulton2015three}
Fulton, B.~J., Weiss, L.~M., Sinukoff, E., {et~al.} 2015, The Astrophysical
  Journal, 805, 175

\bibitem[{Gardner {et~al.}(2006)Gardner, Mather, Clampin, Doyon, Greenhouse,
  Hammel, Hutchings, Jakobsen, Lilly, Long, {et~al.}}]{gardner2006james}
Gardner, J.~P., Mather, J.~C., Clampin, M., {et~al.} 2006, Space Science
  Reviews, 123, 485

\bibitem[{Gelman {et~al.}(2004)Gelman, Carlin, Stern, \&
  Rubin}]{gelman2004bayesian}
Gelman, A., Carlin, J.~B., Stern, H.~S., \& Rubin, D.~B. 2004, CRC Texts in
  Statistical Science

\bibitem[{Gelman {et~al.}(1992)Gelman, Rubin, {et~al.}}]{gelman1992inference}
Gelman, A., Rubin, D.~B., {et~al.} 1992, Statistical science, 7, 457

\bibitem[{Ginzburg {et~al.}(2016)Ginzburg, Schlichting, \&
  Sari}]{ginzburg2016super}
Ginzburg, S., Schlichting, H.~E., \& Sari, R. 2016, The Astrophysical Journal,
  825, 29

\bibitem[{Ginzburg {et~al.}(2018)Ginzburg, Schlichting, \&
  Sari}]{ginzburg2018core}
Ginzburg, S., Schlichting, H.~E., \& Sari, R. 2018, Monthly Notices of the
  Royal Astronomical Society, 476, 759

\bibitem[{{Girardi} {et~al.}(2012){Girardi}, {Barbieri}, {Groenewegen},
  {Marigo}, {Bressan}, {Rocha-Pinto}, {Santiago}, {Camargo}, \& {da
  Costa}}]{girardi12}
{Girardi}, L., {Barbieri}, M., {Groenewegen}, M. A.~T., {et~al.} 2012,
  Astrophysics and Space Science Proceedings, 26, 165

\bibitem[{G{\"u}nther {et~al.}(2019)G{\"u}nther, Pozuelos, Dittmann, Dragomir,
  Kane, Daylan, Feinstein, Huang, Morton, Bonfanti,
  {et~al.}}]{gunther2019super}
G{\"u}nther, M.~N., Pozuelos, F.~J., Dittmann, J.~A., {et~al.} 2019, Nature
  Astronomy, 3, 1099

\bibitem[{Hara {et~al.}(2020)Hara, Bouchy, Stalport, Boisse, Rodrigues,
  Delisle, Santerne, Henry, Arnold, Astudillo-Defru, {et~al.}}]{hara2020sophie}
Hara, N., Bouchy, F., Stalport, M., {et~al.} 2020, Astronomy \& Astrophysics,
  636, L6

\bibitem[{Hara {et~al.}(2017)Hara, Bou{\'e}, Laskar, \&
  Correia}]{hara2017radial}
Hara, N.~C., Bou{\'e}, G., Laskar, J., \& Correia, A. 2017, Monthly Notices of
  the Royal Astronomical Society, 464, 1220

\bibitem[{{Hara} {et~al.}(2021){Hara}, {Delisle}, {Unger}, \&
  {Dumusque}}]{hara2021b}
{Hara}, N.~C., {Delisle}, J.-B., {Unger}, N., \& {Dumusque}, X. 2021, arXiv
  e-prints, arXiv:2106.01365

\bibitem[{{Haywood} {et~al.}(2014){Haywood}, {Collier Cameron}, {Queloz},
  {Barros}, {Deleuil}, {Fares}, {Gillon}, {Lanza}, {Lovis}, {Moutou}, {Pepe},
  {Pollacco}, {Santerne}, {S{\'e}gransan}, \& {Unruh}}]{haywood2014}
{Haywood}, R.~D., {Collier Cameron}, A., {Queloz}, D., {et~al.} 2014, \mnras,
  443, 2517

\bibitem[{Hedges {et~al.}(2020)Hedges, Angus, Barentsen, Saunders, Montet, \&
  Gully-Santiago}]{hedges2020systematics}
Hedges, C., Angus, R., Barentsen, G., {et~al.} 2020, arXiv preprint
  arXiv:2012.08972

\bibitem[{Hippke {et~al.}(2019)Hippke, David, Mulders, \&
  Heller}]{hippke2019wotan}
Hippke, M., David, T.~J., Mulders, G.~D., \& Heller, R. 2019, The Astronomical
  Journal, 158, 143

\bibitem[{{Hippke} \& {Heller}(2019)}]{hippke2019optimized}
{Hippke}, M. \& {Heller}, R. 2019, \aap, 623, A39

\bibitem[{Hobson(2019)}]{Hobson2019}
Hobson, M. 2019, phd thesis, P.92

\bibitem[{Hobson {et~al.}(2018)Hobson, D{\'\i}az, Delfosse, Astudillo-Defru,
  Boisse, Bouchy, Bonfils, Forveille, Hara, Arnold,
  {et~al.}}]{hobson2018sophie}
Hobson, M., D{\'\i}az, R.~F., Delfosse, X., {et~al.} 2018, Astronomy \&
  Astrophysics, 618, A103

\bibitem[{{Hormuth} {et~al.}(2008){Hormuth}, {Brandner}, {Hippler}, \&
  {Henning}}]{hormuth08}
{Hormuth}, F., {Brandner}, W., {Hippler}, S., \& {Henning}, T. 2008, Journal of
  Physics Conference Series, 131, 012051

\bibitem[{Howard {et~al.}(2010)Howard, Johnson, Marcy, Fischer, Wright, Bernat,
  Henry, Peek, Isaacson, Apps, {et~al.}}]{howard2010california}
Howard, A.~W., Johnson, J.~A., Marcy, G.~W., {et~al.} 2010, The Astrophysical
  Journal, 721, 1467

\bibitem[{{Huang} {et~al.}(2020{\natexlab{a}}){Huang}, {Vanderburg}, {P{\'a}l},
  {Sha}, {Yu}, {Fong}, {Fausnaugh}, {Shporer}, {Guerrero}, {Vanderspek}, \&
  {Ricker}}]{2020RNAAS...4..204H}
{Huang}, C.~X., {Vanderburg}, A., {P{\'a}l}, A., {et~al.} 2020{\natexlab{a}},
  Research Notes of the American Astronomical Society, 4, 204

\bibitem[{{Huang} {et~al.}(2020{\natexlab{b}}){Huang}, {Vanderburg}, {P{\'a}l},
  {Sha}, {Yu}, {Fong}, {Fausnaugh}, {Shporer}, {Guerrero}, {Vanderspek}, \&
  {Ricker}}]{2020RNAAS...4..206H}
{Huang}, C.~X., {Vanderburg}, A., {P{\'a}l}, A., {et~al.} 2020{\natexlab{b}},
  Research Notes of the American Astronomical Society, 4, 206

\bibitem[{Isaacson \& Fischer(2010)}]{isaacson2010chromospheric}
Isaacson, H. \& Fischer, D. 2010, The Astrophysical Journal, 725, 875

\bibitem[{{Jenkins}(2002)}]{2002ApJ...575..493J}
{Jenkins}, J.~M. 2002, \apj, 575, 493

\bibitem[{{Jenkins} {et~al.}(2010){Jenkins}, {Chandrasekaran}, {McCauliff},
  {Caldwell}, {Tenenbaum}, {Li}, {Klaus}, {Cote}, \&
  {Middour}}]{2010SPIE.7740E..0DJ}
{Jenkins}, J.~M., {Chandrasekaran}, H., {McCauliff}, S.~D., {et~al.} 2010, in
  Society of Photo-Optical Instrumentation Engineers (SPIE) Conference Series,
  Vol. 7740, Software and Cyberinfrastructure for Astronomy, ed. N.~M.
  {Radziwill} \& A.~{Bridger}, 77400D

\bibitem[{Jenkins {et~al.}(2016)Jenkins, Twicken, McCauliff, Campbell,
  Sanderfer, Lung, Mansouri-Samani, Girouard, Tenenbaum, Klaus,
  {et~al.}}]{jenkins2016tess}
Jenkins, J.~M., Twicken, J.~D., McCauliff, S., {et~al.} 2016, in Software and
  Cyberinfrastructure for Astronomy IV, Vol. 9913, International Society for
  Optics and Photonics, 99133E

\bibitem[{Kley \& Nelson(2012)}]{kley2012planet}
Kley, W. \& Nelson, R. 2012, Annual Review of Astronomy and Astrophysics, 50,
  211

\bibitem[{Lecavelier Des~\'Etangs(2007)}]{des2007diagram}
Lecavelier Des~\'Etangs, A. 2007, Astronomy \& Astrophysics, 461, 1185

\bibitem[{{Li} {et~al.}(2019){Li}, {Tenenbaum}, {Twicken}, {Burke}, {Jenkins},
  {Quintana}, {Rowe}, \& {Seader}}]{Li:DVmodelFit2019}
{Li}, J., {Tenenbaum}, P., {Twicken}, J.~D., {et~al.} 2019, \pasp, 131, 024506

\bibitem[{{Lillo-Box} {et~al.}(2012){Lillo-Box}, {Barrado}, \&
  {Bouy}}]{lillo-box12}
{Lillo-Box}, J., {Barrado}, D., \& {Bouy}, H. 2012, \aap, 546, A10

\bibitem[{{Lillo-Box} {et~al.}(2014){Lillo-Box}, {Barrado}, \&
  {Bouy}}]{lillo-box14b}
{Lillo-Box}, J., {Barrado}, D., \& {Bouy}, H. 2014, \aap, 566, A103

\bibitem[{Lindegren {et~al.}(2018)Lindegren, Hern{\'a}ndez, Bombrun, Klioner,
  Bastian, Ramos-Lerate, De~Torres, Steidelm{\"u}ller, Stephenson, Hobbs,
  {et~al.}}]{lindegren2018gaia}
Lindegren, L., Hern{\'a}ndez, J., Bombrun, A., {et~al.} 2018, Astronomy \&
  Astrophysics, 616, A2

\bibitem[{Lopez \& Fortney(2014)}]{lopez2014understanding}
Lopez, E.~D. \& Fortney, J.~J. 2014, The Astrophysical Journal, 792, 1

\bibitem[{Luque {et~al.}(2018)Luque, Nowak, Pall{\'e}, Dai, Kaminski, Nagel,
  Hidalgo, Bauer, Lafarga, Livingston, {et~al.}}]{luque2018detection}
Luque, R., Nowak, G., Pall{\'e}, E., {et~al.} 2018, arXiv preprint
  arXiv:1812.09242

\bibitem[{Mamajek \& Hillenbrand(2008)}]{mamajek2008improved}
Mamajek, E.~E. \& Hillenbrand, L.~A. 2008, The Astrophysical Journal, 687, 1264

\bibitem[{Mandel \& Agol(2002)}]{mandel2002analytic}
Mandel, K. \& Agol, E. 2002, The Astrophysical Journal Letters, 580, L171

\bibitem[{Martin {et~al.}(2021)Martin, El-Badry, Hod{\v{z}}i{\'c}, Triaud,
  Angus, Birky, Foreman-Mackey, Hedges, Montet, Murphy,
  {et~al.}}]{martin2021toi}
Martin, D.~V., El-Badry, K., Hod{\v{z}}i{\'c}, V.~K., {et~al.} 2021, arXiv
  preprint arXiv:2101.02707

\bibitem[{Mayo {et~al.}(2019)Mayo, Rajpaul, Buchhave, Dressing, Mortier, Zeng,
  Fortenbach, Aigrain, Bonomo, Cameron, {et~al.}}]{mayo201911}
Mayo, A.~W., Rajpaul, V.~M., Buchhave, L.~A., {et~al.} 2019, The Astronomical
  Journal, 158, 165

\bibitem[{{Mazevet} {et~al.}(2019){Mazevet}, {Licari}, {Chabrier}, \&
  {Potekhin}}]{mazevet19}
{Mazevet}, S., {Licari}, A., {Chabrier}, G., \& {Potekhin}, A.~Y. 2019, \aap,
  621, A128

\bibitem[{McLaughlin(1924)}]{mclaughlin1924some}
McLaughlin, D. 1924, The Astrophysical Journal, 60

\bibitem[{McNeil \& Nelson(2010)}]{mcneil2010formation}
McNeil, D. \& Nelson, R. 2010, Monthly Notices of the Royal Astronomical
  Society, 401, 1691

\bibitem[{McQuillan {et~al.}(2013)McQuillan, Aigrain, \&
  Mazeh}]{mcquillan2013measuring}
McQuillan, A., Aigrain, S., \& Mazeh, T. 2013, Monthly Notices of the Royal
  Astronomical Society, 432, 1203

\bibitem[{{Mink}(2011)}]{TRES}
{Mink}, D.~J. 2011, in Astronomical Society of the Pacific Conference Series,
  Vol. 442, Astronomical Data Analysis Software and Systems XX, ed. I.~N.
  {Evans}, A.~{Accomazzi}, D.~J. {Mink}, \& A.~H. {Rots}, 305

\bibitem[{Mortier {et~al.}(2018)Mortier, Bonomo, Rajpaul, Buchhave, Vanderburg,
  Zeng, L{\'o}pez-Morales, Malavolta, Cameron, Dressing,
  {et~al.}}]{mortier2018k2}
Mortier, A., Bonomo, A., Rajpaul, V., {et~al.} 2018, arXiv preprint
  arXiv:1808.08187

\bibitem[{Mousis {et~al.}(2020)Mousis, Deleuil, Aguichine, Marcq, Naar,
  Aguirre, Brugger, \& Gon{\c{c}}alves}]{mousis20}
Mousis, O., Deleuil, M., Aguichine, A., {et~al.} 2020, The Astrophysical
  Journal, 896, L22

\bibitem[{Nielsen {et~al.}(2020)Nielsen, Gandolfi, Armstrong, Jenkins,
  Fridlund, Santos, Dai, Adibekyan, Luque, Steffen, {et~al.}}]{nielsen2020mass}
Nielsen, L.~D., Gandolfi, D., Armstrong, D., {et~al.} 2020, Monthly Notices of
  the Royal Astronomical Society, 492, 5399

\bibitem[{Noyes {et~al.}(1984)Noyes, Hartmann, Baliunas, Duncan, \&
  Vaughan}]{noyes1984rotation}
Noyes, R., Hartmann, L., Baliunas, S., Duncan, D., \& Vaughan, A. 1984, The
  Astrophysical Journal, 279, 763

\bibitem[{Osborn {et~al.}(2017)Osborn, Santerne, Barros, Santos, Dumusque,
  Malavolta, Armstrong, Hojjatpanah, Demangeon, Adibekyan,
  {et~al.}}]{osborn2017k2}
Osborn, H., Santerne, A., Barros, S., {et~al.} 2017, Astronomy \& Astrophysics,
  604, A19

\bibitem[{Otegi {et~al.}(2020)Otegi, Bouchy, \& Helled}]{otegi2020revisited}
Otegi, J., Bouchy, F., \& Helled, R. 2020, Astronomy \& Astrophysics, 634, A43

\bibitem[{Owen \& Wu(2013)}]{owen2013kepler}
Owen, J.~E. \& Wu, Y. 2013, The Astrophysical Journal, 775, 105

\bibitem[{Owen \& Wu(2017)}]{owen2017evaporation}
Owen, J.~E. \& Wu, Y. 2017, The Astrophysical Journal, 847, 29

\bibitem[{Paulson {et~al.}(2002)Paulson, Saar, Cochran, \&
  Hatzes}]{paulson2002searching}
Paulson, D.~B., Saar, S.~H., Cochran, W.~D., \& Hatzes, A.~P. 2002, The
  Astronomical Journal, 124, 572

\bibitem[{Pedregosa {et~al.}(2011)Pedregosa, Varoquaux, Gramfort, Michel,
  Thirion, Grisel, Blondel, Prettenhofer, Weiss, Dubourg, Vanderplas, Passos,
  Cournapeau, Brucher, Perrot, \& {{\'E}}douard Duchesnay}]{Pedregosa-11}
Pedregosa, F., Varoquaux, G., Gramfort, A., {et~al.} 2011, Journal of Machine
  Learning Research, 12, 2825

\bibitem[{Pepe {et~al.}(2002)Pepe, Mayor, Galland, Naef, Queloz, Santos, Udry,
  \& Burnet}]{pepe2002coralie}
Pepe, F., Mayor, M., Galland, F., {et~al.} 2002, Astronomy \& Astrophysics,
  388, 632

\bibitem[{Perruchot {et~al.}(2008)Perruchot, Kohler, Bouchy, Richaud, Richaud,
  Moreaux, Merzougui, Sottile, Hill, Knispel, {et~al.}}]{perruchot2008sophie}
Perruchot, S., Kohler, D., Bouchy, F., {et~al.} 2008, 7014, 70140J

\bibitem[{Ricker {et~al.}(2015)Ricker, Winn, Vanderspek,
  {et~al.}}]{ricker2015jatis}
Ricker, G., Winn, J., Vanderspek, R., {et~al.} 2015, JATIS, 1, 014003

\bibitem[{Rogers(2015)}]{rogers2015most}
Rogers, L.~A. 2015, The Astrophysical Journal, 801, 41

\bibitem[{Rossiter(1924)}]{rossiter1924detection}
Rossiter, R. 1924, The Astrophysical Journal, 60

\bibitem[{Santerne {et~al.}(2012)Santerne, D{\'\i}az, Moutou, Bouchy,
  H{\'e}brard, Almenara, Bonomo, Deleuil, \& Santos}]{santerne2012sophie}
Santerne, A., D{\'\i}az, R., Moutou, C., {et~al.} 2012, Astronomy \&
  Astrophysics, 545, A76

\bibitem[{Santos {et~al.}(2013)Santos, Sousa, Mortier, Neves, Adibekyan,
  Tsantaki, Mena, Bonfils, Israelian, Mayor, {et~al.}}]{santos2013sweet}
Santos, N., Sousa, S., Mortier, A., {et~al.} 2013, Astronomy \& Astrophysics,
  556, A150

\bibitem[{Schlafly \& Finkbeiner(2011)}]{schlafly2011measuring}
Schlafly, E.~F. \& Finkbeiner, D.~P. 2011, The Astrophysical Journal, 737, 103

\bibitem[{Schlecker {et~al.}(2020)Schlecker, Mordasini, Emsenhuber, Klahr,
  Henning, \& Burn}]{schlecker2020new}
Schlecker, M., Mordasini, C., Emsenhuber, A., {et~al.} 2020, Astronomy \&
  Astrophysics

\bibitem[{Schlegel {et~al.}(1998)Schlegel, Finkbeiner, \&
  Davis}]{schlegel1998maps}
Schlegel, D.~J., Finkbeiner, D.~P., \& Davis, M. 1998, The Astrophysical
  Journal, 500, 525

\bibitem[{Schlichting(2014)}]{schlichting2014formation}
Schlichting, H.~E. 2014, The Astrophysical Journal Letters, 795, L15

\bibitem[{Schoenberg(1946)}]{schoenberg1946contributions}
Schoenberg, I.~J. 1946, Quarterly of Applied Mathematics, 4, 112

\bibitem[{Smith {et~al.}(2012)Smith, Stumpe, Van~Cleve, Jenkins, Barclay,
  Fanelli, Girouard, Kolodziejczak, McCauliff, Morris,
  {et~al.}}]{smith2012kepler}
Smith, J.~C., Stumpe, M.~C., Van~Cleve, J.~E., {et~al.} 2012, Publications of
  the Astronomical Society of the Pacific, 124, 1000

\bibitem[{{Sotin} {et~al.}(2007){Sotin}, {Grasset}, \& {Mocquet}}]{sotin07}
{Sotin}, C., {Grasset}, O., \& {Mocquet}, A. 2007, \icarus, 191, 337

\bibitem[{Sousa {et~al.}(2018)Sousa, Adibekyan, Delgado-Mena, Santos,
  Andreasen, Ferreira, Tsantaki, Barros, Demangeon, Israelian,
  {et~al.}}]{sousa2018sweet}
Sousa, S., Adibekyan, V., Delgado-Mena, E., {et~al.} 2018, Astronomy \&
  Astrophysics, 620, A58

\bibitem[{Stassun {et~al.}(2017{\natexlab{a}})Stassun, Collins, \&
  Gaudi}]{stassun2017accurate}
Stassun, K.~G., Collins, K.~A., \& Gaudi, B.~S. 2017{\natexlab{a}}, The
  Astronomical Journal, 153, 136

\bibitem[{Stassun {et~al.}(2017{\natexlab{b}})Stassun, Corsaro, Pepper, \&
  Gaudi}]{stassun2017empirical}
Stassun, K.~G., Corsaro, E., Pepper, J.~A., \& Gaudi, B.~S. 2017{\natexlab{b}},
  The Astronomical Journal, 155, 22

\bibitem[{Stassun \& Torres(2016)}]{stassun2016eclipsing}
Stassun, K.~G. \& Torres, G. 2016, The Astronomical Journal, 152, 180

\bibitem[{Stassun \& Torres(2021)}]{stassun2021parallax}
Stassun, K.~G. \& Torres, G. 2021, arXiv preprint arXiv:2101.03425

\bibitem[{{Strehl}(1902)}]{strehl1902}
{Strehl}, K. 1902, Astronomische Nachrichten, 158, 89

\bibitem[{Stumpe {et~al.}(2014)Stumpe, Smith, Catanzarite, Van~Cleve, Jenkins,
  Twicken, \& Girouard}]{stumpe2014multiscale}
Stumpe, M.~C., Smith, J.~C., Catanzarite, J.~H., {et~al.} 2014, Publications of
  the Astronomical Society of the Pacific, 126, 100

\bibitem[{{Stumpe} {et~al.}(2012){Stumpe}, {Smith}, {Van Cleve}, {Twicken},
  {Barclay}, {Fanelli}, {Girouard}, {Jenkins}, {Kolodziejczak}, {McCauliff}, \&
  {Morris}}]{Stumpe2012}
{Stumpe}, M.~C., {Smith}, J.~C., {Van Cleve}, J.~E., {et~al.} 2012, \pasp, 124,
  985

\bibitem[{Sun {et~al.}(2019)Sun, Ioannidis, Gu, Schmitt, Wang, \&
  Kouwenhoven}]{sun2019kepler}
Sun, L., Ioannidis, P., Gu, S., {et~al.} 2019, Astronomy \& Astrophysics, 624,
  A15

\bibitem[{{Telting} {et~al.}(2014){Telting}, {Avila}, {Buchhave}, {Frandsen},
  {Gandolfi}, {Lindberg}, {Stempels}, {Prins}, \& {NOT staff}}]{Telting2014}
{Telting}, J.~H., {Avila}, G., {Buchhave}, L., {et~al.} 2014, Astronomische
  Nachrichten, 335, 41

\bibitem[{Torres {et~al.}(2010)Torres, Andersen, \&
  Gim{\'e}nez}]{torres2010accurate}
Torres, G., Andersen, J., \& Gim{\'e}nez, A. 2010, The Astronomy and
  Astrophysics Review, 18, 67

\bibitem[{{Twicken} {et~al.}(2018){Twicken}, {Catanzarite}, {Clarke},
  {Girouard}, {Jenkins}, {Klaus}, {Li}, {McCauliff}, {Seader}, {Tenenbaum},
  {Wohler}, {Bryson}, {Burke}, {Caldwell}, {Haas}, {Henze}, \&
  {Sanderfer}}]{Twicken:DVdiagnostics2018}
{Twicken}, J.~D., {Catanzarite}, J.~H., {Clarke}, B.~D., {et~al.} 2018, \pasp,
  130, 064502

\bibitem[{Zechmeister {et~al.}(2018)Zechmeister, Reiners, Amado, Azzaro, Bauer,
  B{\'e}jar, Caballero, Guenther, Hagen, Jeffers,
  {et~al.}}]{zechmeister2018spectrum}
Zechmeister, M., Reiners, A., Amado, P.~J., {et~al.} 2018, Astronomy \&
  Astrophysics, 609, A12

\bibitem[{Zhu \& Wu(2018)}]{zhu2018super}
Zhu, W. \& Wu, Y. 2018, The Astronomical Journal, 156, 92

\end{thebibliography}
%%%%%%%%%%%%%%%%%%%%%%%%%%%%%%%%%%%%%%%%%%%%%%%%%%%%%%%%%%%%%%
\begin{appendix}

\onecolumn
\section{RV's time series}
\begin{table}[h]\centering

\caption{SOPHIE RVs for HD207897 b}
\label{tab:rvs sophie}
\resizebox{0.8\columnwidth}{!}{%
\begin{tabular}{c c c c c c c c c}
\hline
\hline
BJD (-2400000 d)&	RV (km s$^{−1}$ )  & $\sigma_{RV}$ (km s$^{−1}$ )& 	FWHM (km s$^{−1}$ )	& BIS (km s$^{−1}$ )& H$\alpha$ & $\sigma_{H\alpha}$ &CRX&$\sigma_{CRX}$ \\
\hline
56117.5531& -6.3203&    0.0015& 6.6803& -0.0025&0.1294&	0.0005& -39.0617& 13.9761\\
56118.5849&	-6.3205&	0.0015&	6.6846&	0.0027&	0.1305&	0.0005& -34.4212& 12.8526  \\
56119.5401&	-6.3238&	0.0016&	6.6849&	-0.0016&0.1296&	0.0006& -12.6147& 12.0847\\
56120.5123&	-6.3206&	0.0016&	6.6712&	-0.0006& 0.1306&0.0006& -29.3660& 11.2578 \\
56137.5869&	-6.3240&	0.0016&	6.7125&	0.0011& 0.1323&	0.0006 &-36.7534& 14.7557 \\
56138.5331&	-6.3249&	0.0017&	6.6945&	0.0065& 0.1335&	0.0007  &-28.7203& 14.5124\\
56139.5731&	-6.3300&	0.0015&	6.6839&	-0.0008&0.1296&	0.0005  -&32.8700& 12.5074\\
56140.5330&	-6.3310&	0.0016&	6.6906&	0.0043& 0.1310&	0.0006 &-34.3580& 10.0305 \\
56150.5049&	-6.3162&	0.0015&	6.6867&	-0.0028& 0.1343&0.0005 &-7.1528& 14.0474 \\
56152.5655&	-6.3214&	0.0018&	6.6849&	0.0002& 0.1334&	0.0008 &-9.6228& 14.1995\\
56167.5191&	-6.3227&	0.0015&	6.6823&	0.0054&	0.1302&	0.0006 &-53.2894& 10.1835 \\
56178.3748&	-6.3278&	0.0015&	6.7033&	0.0013&  0.1323&0.0006 &-46.0315& 12.8783 \\
56497.5314&	-6.3279&	0.0018&	6.6991&	-0.0020& 0.1288&0.0008 &-27.5805& 15.6965 \\
56498.5527&	-6.3276&	0.0016&	6.7072&	0.0070&	0.1326&	0.0006  &-6.9804& 14.2738\\
56499.5446&	-6.3316&	0.0016&	6.7074&	-0.0023&0.1316&	0.0006 &-41.0561& 13.8978\\
56500.5510&	-6.3333&	0.0017&	6.6891&	-0.0059&0.1321&	0.0007 &5.3262& 13.2646 \\
56517.4678&	-6.3314&	0.0016&	6.7181&	0.0051&	0.1337&	0.0007 &15.8414& 9.3975 \\
56519.4789&	-6.3304&	0.0016&	6.7067&	-0.0048&0.1332&	0.0007 &-18.4473& 11.5302\\
56520.4837&	-6.3319&	0.0016&	6.7042&	-0.0029	&0.1345&0.0006 &20.1748& 12.7335 \\
56521.4875&	-6.3261&	0.0016&	6.6994&	-0.0011 &0.1338&0.0007 &-12.2005 &14.9481\\
56522.4818&	-6.3248&	0.0016&	6.7130&	0.0057	&0.1331&	0.0006 &-9.6604& 11.5639  \\
56523.4905&	-6.3274&	0.0015&	6.7106&	0.0077	&0.1319&	0.0006 &-36.2878& 13.6435\\
56524.4789&	-6.3246&	0.0024&	6.6754&	0.0101  &0.1384&	0.0013  &-0.9706& 21.2910\\
56558.3761&	-6.3225&	0.0015&	6.7002&	-0.0020	&0.1301&	0.0006 &-12.9535& 15.0095\\
56560.3978&	-6.3254&	0.0016&	6.7038&	0.0003  &0.1311&	0.0006 &-30.1229& 14.8829\\
56583.3673&	-6.3305&	0.0017&	6.6867&	-0.0021	&0.1307&	0.0008 &-6.3707& 16.6035\\
56592.3536&	-6.3289&	0.0021&	6.6841&	-0.0077	&0.1321&	0.0011 &-8.3097& 21.0877\\
56624.3190&	-6.3292&	0.0019&	6.7214&	0.0091	&0.1346&	0.0009 &6.39867& 17.5343 \\
56625.2867&	-6.3321&	0.0015&	6.7172&	-0.0040	&0.1310&	0.0006 &17.7943& 10.9701\\
56626.2718&	-6.3297&	0.0015&	6.7156&	0.0024	&0.1327&	0.0006 &-3.3048& 12.3435\\
56628.3163&	-6.3305&	0.0016&	6.7214&	0.0007  &0.1322&	0.0007 &-8.6098& 14.9817\\
56629.3041&	-6.3354&	0.0017&	6.7138&	-0.0064	&0.1322&	0.0007 &11.1813& 13.8973\\
56630.2650&	-6.3339&	0.0014&	6.7173&	0.0018  &0.1304&	0.0005 &-11.7681& 12.1465\\
56631.2725&	-6.3342&	0.0014&	6.7182&	-0.0030	&0.1297&	0.0005 &-23.8070& 12.1372 \\
56654.2301&	-6.3252&	0.0019&	6.7088& -0.0011	&0.1349&	0.0009  &-15.5221& 17.2095\\
56656.2313&	-6.3296&	0.0018&	6.7171&	0.0054	&0.1324&	0.0008 &2.0870& 14.7556 \\
56657.2325&	-6.3260&	0.0015&	6.7183& 0.0059	&0.1325&	0.0006 &21.1445& 11.7855\\
56664.2359&	-6.3336&	0.0023&	6.7235&	0.0026	&0.1303&	0.0013 &-9.4066& 21.0183\\
57214.5707&	-6.3257&	0.0013&	6.7541&	-0.0027	&0.1291&	0.0005 &4.0133& 13.6434\\
57217.5644&	-6.3202&	0.0013&	6.7651&	0.0025  &0.1284&	0.0006 &5.5657& 16.3052 \\
57284.4185&	-6.3216&	0.0016&	6.7057&	-0.0038	&0.1293&	0.0007 &-3.4004& 12.3004 \\
58862.3082&	-6.3280&	0.0018&	6.7636&	0.0076	&0.1415&	0.0009 &-1.5094& 21.4660\\
58875.3024&	-6.3284&	0.0016&	6.8431&	-0.0004	&0.1440&	0.0007 &108.6779& 21.7024\\
58877.3789&	-6.3325&	0.0023&	6.7005&	0.0187  &0.1470&	0.0013 &69.2134& 29.4011\\
58878.3249&	-6.3292&	0.0019&	6.7894&	-0.000  &0.1437&	0.0009  &57.8937& 28.6864\\
58879.3132&	-6.3368&	0.0022&	6.8407&	-0.0178	&0.1444&	0.0011 &54.4243& 36.3337\\
58880.2542&	-6.3327&	0.0012&	6.8466&	0.0038	&0.1414&	0.0005 &53.1909& 18.2402\\
58881.2536&	-6.3336&	0.0016&	6.7832&	-0.0118	&0.1455&	0.0007 &49.2682& 15.8376\\
58882.6853&	-6.3330&	0.0025&	6.7353&	-0.0007	&0.1445&	0.0014 &80.9618& 43.6620\\
58883.2559&	-6.3304&	0.0017&	6.7916&	-0.0008 &0.1442&	0.0008 &60.3972& 18.5206\\
58885.2949&	-6.3384&	0.0020&	6.7925&	-0.0120	&0.1415&	0.0010 &102.8511& 29.6527\\
58886.2969&	-6.3374&	0.0017&	6.8429&	-0.0090	&0.1438&	0.0008 &50.0756& 21.1131\\
58888.2703&	-6.3319&	0.0016&	6.8534&	-0.0097	&0.1424&	0.0007 &84.0189& 25.1804\\
58890.2920&	-6.3314&	0.0023&	6.7950&	-0.0084	&0.1472&	0.0012 &70.5820& 30.6721\\
58892.2593&	-6.3200&	0.0023&	6.7914&	0.0029	&0.1485&	0.0012 &61.7681& 27.8826\\
58894.2567&	-6.3309&	0.0014&	6.8221&	-0.0060 &0.1436&	0.0006 &85.0683& 18.1804\\
58895.2570&	-6.3419&	0.0017&	6.8537&	0.0008	&0.1447&	0.0008 &76.4210& 22.0550\\
58898.2952&	-6.3335&	0.0021&	6.7098&	0.0081  &0.1446&	0.0012 &-12.7695& 29.9844\\
58899.2728&	-6.3337&	0.0019&	6.7510&	-0.0144	&0.1441&	0.0009 &-4.7509& 23.2686\\
58900.2705&	-6.3272&	0.0013&	6.8566&	-0.0026	&0.1399&	0.0005 &17.5256& 13.1316\\
58901.2780&	-6.3219&	0.0018&	6.8054&	0.0090	&0.1453&	0.0009 &6.4814& 19.9266\\
58902.2708&	-6.3301&	0.0012&	6.8455&	-0.0022	&0.1419&	0.0005& 21.4611& 14.2376\\
58903.2734&	-6.3223&	0.0016&	6.8011&	-0.0038	&0.1415&	0.0007& 67.6199& 20.3261\\
59068.5491&	-6.3265&	0.0011&	6.6780&	0.0011  &----&      -----& -------& ------	\\					

\hline
\end{tabular}
    }

\end{table}	

\begin{table}[h]
\centering

\caption{HIRES RVs for HD207897 b}
\label{rvs hires}
\resizebox{0.5\columnwidth}{!}{%
\begin{tabular}{l c c c c }
\hline
\hline
BJD (-2400000 d)&	RV (m s$^{−1}$ ) & $\sigma_{RV}$(m s$^{−1}$ ) & S-index& $\sigma_{S-index}$\\
\hline
52832.039671&	-6.1499&	1.04&---&----\\
52855.027852&	-6.4812&	2.42&---&----\\
53196.053218&	-8.9158&	1.01&---&----\\
\hline
53602.917000&	1.127055&	0.819542&	0.231000&	0.001000\\
53962.008000&	4.990706&	0.870220&	0.222000&	0.001000\\
56588.935000&	3.431498&	1.463105&	0.225000&	0.001000\\
57237.106000&	3.290118&	1.190904&	0.215400&	0.001000\\
58869.707000&	3.595136&	1.360255&	0.249000&	0.001000\\
59004.080000&	-3.829985&	1.708751&	0.251100&	0.001000\\
59007.977000&	0.107424&	1.636860&	0.247300&	0.001000\\
59008.017000&	0.038845&	1.893359&	0.249900&	0.001000\\
59008.066000&	-3.912803&	1.738616&	0.250000&	0.001000\\
59039.084000&	-2.874458&	1.880687&	0.240000&	0.001000\\
59079.039000&	3.561938&	1.538733&	0.238600&	0.001000\\
59089.002000&	0.991416&	1.012931&	0.233500&	0.001000\\
59090.031000&	-1.806837&	0.977288&	0.236100&	0.001000\\
59091.056000&	-3.314227&	0.904606&	0.233400&	0.001000\\
59092.060000&	-3.337441&	0.963977&	0.237100&	0.001000\\
59093.027000&	-0.185500&	0.992191&	0.239300&	0.001000\\
59094.984000&	-4.733336&	0.983995&	0.233400&	0.001000\\
59097.973000&	2.455004&	1.046085&	0.235400&	0.001000\\
59099.827000&	2.774576&	1.035973&	0.236600&	0.001000\\
59100.985000&	4.753109&	1.043153&	0.237800&	0.001000\\
59102.009000&	3.388690&	0.992955&	0.237100&	0.001000\\
59115.059000&	8.057584&	0.990576&	0.244300&	0.001000\\
59118.985000&	0.615674&	1.028696&	0.240700&	0.001000\\
59119.868000&	4.070617&	1.021998&	0.239300&	0.001000\\
59120.980000&	-0.036780&	0.993603&	0.241600&	0.001000\\
59122.955000&	-1.142312&	0.922378&	0.239500&	0.001000\\
59123.937000&	-2.279427&	0.982050&	0.238200&	0.001000\\
59142.961000&	-4.944806&	1.161786&	0.242200&	0.001000\\
59153.887000&	-2.863129&	1.141554&	0.240700&	0.001000\\
59181.890000&	-1.681638&	1.441832&	0.236500&	0.001000\\
59187.825000&	-1.590208&	1.406428&	0.227200&	0.001000\\
59188.813000&	-6.084715&	1.833509&	0.227800&	0.001000\\
59189.870000&	-6.143414&	1.421509&	0.229800&	0.001000\\

\hline
    \end{tabular}
    }
\tablefoot{The first three data were taken with a different CCD detector. For the sake of simplicity, we excluded them in our joint modeling by EXOFASTv2.}
    
\end{table}	

\begin{table}[h]\centering

\caption{APF RVs for HD207897 b}
\label{tab:rvs apf}
\resizebox{0.4\columnwidth}{!}{%
  \begin{tabular}{l c c c}
\hline
\hline
BJD (-2400000 d)&	RV (m s$^{−1}$ ) & $\sigma_{RV}$(m s$^{−1}$ )\\
\hline
59002.885&	3.030364933&	3.218557358\\					
59003.809&	5.987553939&	2.861260653\\					
59004.916&	4.500414023&	2.633646965\\				
59005.881&	2.367894193&	2.662182808\\					
59006.804&	14.66981698&	4.979400635\\					
59008.864&	1.427995591&	3.299036264\\					
59009.826&	-1.594891397&	2.404323578\\					
59010.938&	-0.297586458&	2.476657391\\			
59011.895&	-5.612035257&	2.451689482\\				
59012.812&	3.174356114&	4.957623482\\				
59016.811&	11.1418913&	    3.183858633\\					
59017.938&	9.565100481&	3.032974005\\					
59018.847&	8.944296078&	2.887774467\\				
59039.912&	-1.683343659&	2.742200613\\					
59059.929&	-5.287302366&	2.546239138\\				
59130.705&	8.958821818&	3.813406229\\					
59151.657&	-1.77725919&	2.748860359\\					
59178.67&	-3.064406035&	2.962447405\\									
59203.816&	-1.276284674&	3.144914389\\					
59223.847&	-3.812726288&	3.542819262\\					
59252.786&	-4.278596616&	4.021532536\\					
59273.802&	-3.184576312&	3.384922504\\

\hline
 \end{tabular}%
}
\end{table}

\onecolumn
\section{Figures}

\begin{figure}[h!]
\centering
 	\includegraphics[width=0.5\columnwidth]{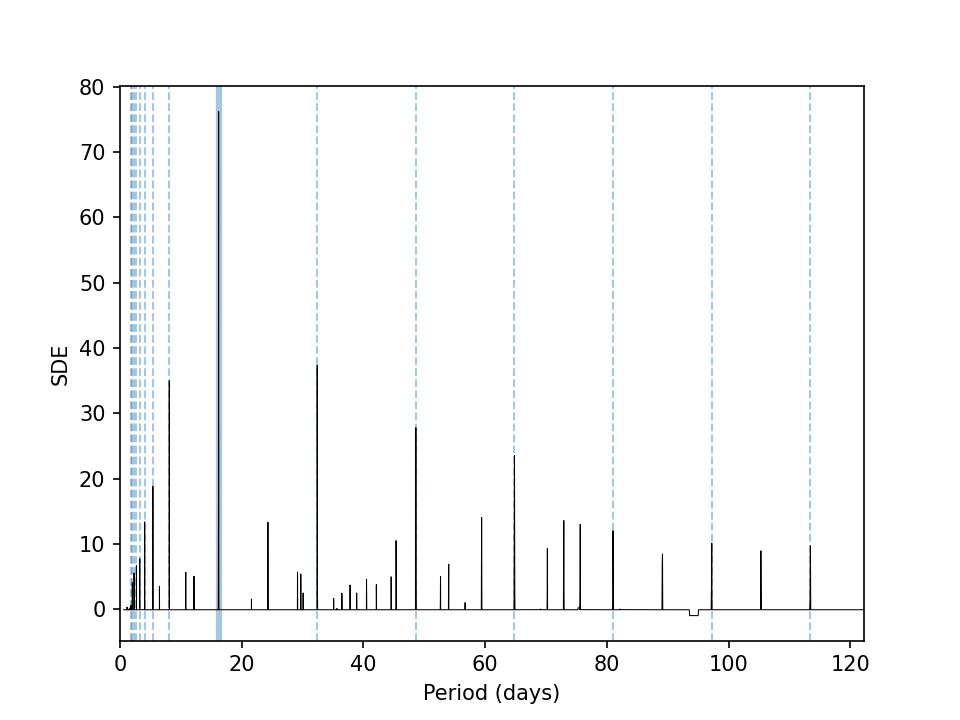}
	\caption{TLS periodogram for the TESS light curve of HD207897 b. The blue line shows the highest peak at 16.20 d with SDE=76.2. The dashed blue lines indicate the aliases of this period.}
	\label{fig:tls-sed}
\end{figure}

\begin{figure}[h!]
\centering
\includegraphics[width=0.5\columnwidth]{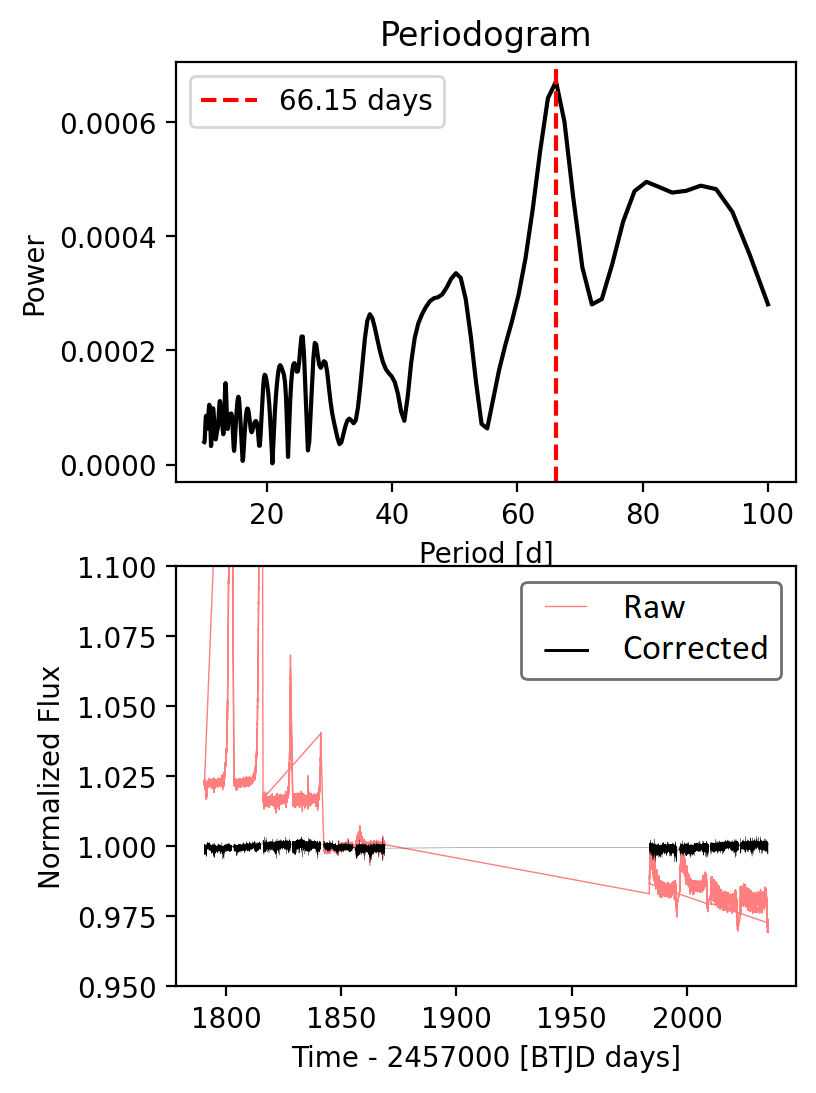}
\caption{\emph{Top}: Systematics-Insensitive Periodogram (SIP) for HD207897 b. \emph{Bottom}: raw SAP TESS light curves in red and detrended data against instrument systematics by SIP in black. The SIP periodogram does not exhibit any significant signal for HD207897 b, since the highest peak in this periodogram has a power of only 0.0006 at 66.6 d. }
\label{fig:SIP}
\end{figure}

\end{appendix}

\end{document}